\documentclass[aps,pra,reprint,groupedaddress]{revtex4-2}
\usepackage{amsmath}
\usepackage{amsfonts}
\usepackage{amssymb}
\usepackage{graphicx}
\usepackage{empheq}
\usepackage{comment}
\usepackage{diagbox}
\usepackage{orcidlink}
\usepackage{tikz}
\usepackage{xcolor}
\usepackage{lipsum}
\usepackage{hyperref}
\usepackage{gensymb}
\definecolor{linkcolour}{HTML}{000066}	
\hypersetup{colorlinks,breaklinks,
	urlcolor=linkcolour, 
	linkcolor=linkcolour,
	citecolor=linkcolour}

\hypersetup{colorlinks=true, urlcolor=blue, linkcolor=blue, citecolor=blue,plainpages=false}

\newcommand{\dee}{\mathrm{d}}
\newcommand{\red}[1]{#1}

\newenvironment{redtext}
{}
{}

\begin{document}

\preprint{APS/123-QED}

\title{Extended scattering channels for random matrix simulations of polarized light transport} 

\author{Niall Byrnes$^1$\orcidlink{0000-0002-1554-3820}, Sulagna Dutta$^{2}$\orcidlink{0000-0003-4361-8540}, and Matthew R. Foreman$^{1,2}$\orcidlink{0000-0001-5864-9636}}
\email[]{matthew.foreman@ntu.edu.sg}
\affiliation{
$^1$School of Electrical and Electronic Engineering, Nanyang Technological University, 50 Nanyang Avenue, Singapore 639798 \\
$^2$Institute for Digital Molecular Analytics and Science, 59 Nanyang Drive, Singapore 636921
}

\date{\today}

\begin{abstract}
Modeling the propagation of light through disordered media is central to understanding and controlling wave transport in diverse optical and mesoscopic applications. Here, we present a random matrix simulation framework for modeling the transport of polarized light through random media composed of arbitrary particulate scatterers. Our approach employs extended scattering channels applied to angular spectral decompositions of the underlying fields, enabling flexible representations of arbitrary illumination and detection profiles. In contrast to previous work, this framework provides a rigorous treatment of scattering matrix correlations and offers \red{a novel geometric approach to computing tilt memory effect correlations.} We provide a detailed exposition of the underlying theory and illustrate several key features through numerical simulations. Our work is supported by a free accompanying codebase.\end{abstract}

\maketitle

\section{Introduction}
Modeling wave propagation through disordered media is a central challenge across  many areas of science and technology, including imaging \cite{Mosk2012}, remote sensing \cite{Li2022}, wireless communications \cite{ALOZIE2023e01816}, underwater acoustics \cite{6605638}, and metasurfaces \cite{Lalanne:25}. Many interesting physical phenomena pertinent to these fields emerge in the mesoscopic regime, where transport properties are determined by the interference of propagating modes traversing many complex scattering paths through the system~\cite{RevModPhys.89.015005}. A central tool in many theoretical and experimental studies of mesoscopic wave transport is the scattering matrix formalism, which, in the optical domain, most generally describes the linear mapping between the amplitude, phase and polarization state of electromagnetic waves incident upon and emerging from a system \cite{Tripathi:12}. Due to this generality, the scattering matrix is highly versatile and has therefore seen extensive application in optics including in focusing through scattering media~\cite{Vellekoop:07}, enhancing transmission~\cite{Kim2012}, depth-targeted energy deposition~\cite{Bender2022}, crafting maximum information optical states~\cite{Bouchet2021}, enhancing sensitivity to perturbations~\cite{GutirrezCuevas2024}, laser cooling~\cite{Hpfl2023}, quantifying Fisher information~\cite{Hpfl2024}, describing resonance perturbations in complex perturbed systems~\cite{Byrnes2025b, Byrnes2025a, Wang2025b}, and investigating electromagnetic exceptional points~\cite{Shou2025} .

Different realizations of a disordered system that are macroscopically indistinguishable typically possess distinct microscopic configurations. Even subtle differences in these microscopic arrangements can dramatically affect wave transport, resulting in markedly different scattering matrices. As a result, the theoretical calculation of the scattering matrix for a specific realization is generally infeasible in practice. Consequently, statistical approaches based on the generation of symmetry-constrained random matrices have been employed \cite{RevModPhys.69.731, 86th-pl8q}. These methods have revealed universal features of wave transport in complex media and have been applied to a wide range of phenomena, including universal conductance fluctuations~\cite{PhysRevLett.55.1622}, coherent backscattering~\cite{wolf1985weak}, Anderson localization~\cite{mafi2015transverse}, time-delay statistics~\cite{Fyodorov1997} and fundamental limits on information transmission~\cite{Byrnes2020njp}. The universality of classic random matrix models, however, also limits their ability to capture the diverse phenomena that arise in different classes of systems throughout nature.

In recent works, we introduced an approach~\cite{Byrnes14122022} and accompanying codebase~\cite{oldrmt} for generating random scattering matrices that made two key advancements over previous schemes. Firstly, we presented a framework general enough to describe media composed of arbitrary particulate scatterers, thereby allowing modelling of more realistic media whilst still enabling studies of broad classes of systems to identify universal trends. Secondly, we explicitly incorporated, for the first time, the polarization properties of light meaningfully into a random matrix model. Polarization is frequently neglected in scattering matrix studies, as the relevant theory is often simpler under a scalar approximation. While this is often justified on the basis of multiple scattering induced depolarization, polarization nevertheless remains significant at intermediate scales and is therefore important for accurately capturing wave propagation in such mesoscopic regimes~\cite{10.3389/fphy.2022.815296, BYRNES2022127462}. Moreover, the rate of depolarization is not universal and can be influenced by the constituent scatterers in the medium and by the incident field's polarization state~\cite{PhysRevB.40.9342}. Polarization also remains relevant in reflection-mode studies, for example when probing targets located near the surface of thick scattering medium~\cite{10.1117/1.JBO.21.7.071107}. 

Whilst the approach of Ref.~\cite{Byrnes14122022} successfully describes a number of polarization related phenomena, it nevertheless suffers from several limitations. Most notably, physical memory effect correlations, which can play an important role in imaging~\cite{Yang2014} and cryptographic~\cite{Liao2017} applications, were absent from generated scattering matrices. More technically, the normalization of the Dirac delta function appearing in scattering matrix correlations, was also handled in an ad hoc manner. Practically speaking, our previous codebase also employed a discrete, rectangular grid of wavevectors to sample incident electric fields, which does not naturally conform to the circular geometry of Fourier space, nor to many common beam profiles. In this work, we therefore introduce a new method for generating random matrices that resolves these issues by representing incident fields using arbitrarily shaped, extended scattering channels in the Fourier (angular spectrum) domain. Point samples of the electric fields are replaced by piecewise averages, leading to a \red{novel geometric approach to calculating memory effect correlations. As shall be demonstrated, our method ultimately allows us to generate random scattering matrices with the memory effect visible at the level of single realizations}. Accompanying this article is a new Python codebase~\cite{rmtnew} utilising the improved framework. It furthermore includes other computational optimizations that enable the calculation of scattering matrices substantially larger than those previously achievable, thereby better capturing the high mode counts in many real world optical systems \cite{YU201533}. 

This paper is organized into two main sections. In Sec.~\ref{sec:theory}, we introduce the relevant theory behind our matrix generation algorithm. We begin with a description of the problem in Sec.~\ref{sec:problem}, and then introduce our extended channel approach in Sec.~\ref{sec:extended}. In Sec.~\ref{sec:statsmodel}, we outline our physical and statistical assumptions for the scattering media under consideration and derive expressions for the mean scattering matrix in Sec.~\ref{sec:mean} and scattering-matrix correlations in Sec.~\ref{sec:corr}. In Sec.~\ref{sec:generation}, we discuss practical considerations and outline our computational method for generating random scattering matrices, \red{before summarizing the algorithmic workflow in Sec.~\ref{sec:summary}}. Finally, in Sec.~\ref{sec:numerical}, we present several numerical examples illustrating the new features of our code. We first model a Hermite-Gaussian beam with extended channels in Sec.~\ref{sec:fields}, demonstrate memory effect correlations in Sec.~\ref{sec:memory}, and finally illustrate the propagation of several beams with different polarization structures in Sec.~\ref{sec:cascade}.

\section{Theory}
\label{sec:theory}
\subsection{Electric field expressions and the continuous scattering matrix}
\label{sec:problem}
We consider a three-dimensional space in which the macroscopic refractive index is unity everywhere except within a slab $\mathcal{S}$ (the scattering medium) of thickness $L$ located at $-L/2 < z < L/2$. Within $\mathcal{S}$, the refractive index varies piecewise continuously over a countable set of compact sub-regions, representing individual scatterers randomly distributed in space. In realistic scenarios, $\mathcal{S}$ may also have a physical boundary at the planes $ z = \pm L/2$, as well as a background refractive index not equal to unity. A time-harmonic electric field $\mathbf{E}(\mathbf{r}, t)$, with optical frequency $\omega$ and complex envelope $\mathbf{E}(\mathbf{r})$ satisfying $\mathbf{E}(\mathbf{r}, t) = \mathrm{Re}[\mathbf{E}(\mathbf{r}) e^{-i\omega t}]$ exists throughout space due to sources of electromagnetic waves located far from $\mathcal{S}$, i.e., at $|z| \gg L/2$.

Let $\mathcal{R}^+$ and $\mathcal{R}^-$ denote the vacuum regions $z > L/2$ and $z < -L/2$ respectively. In either of these regions, the electric field may be represented as a superposition of counter-propagating plane waves~\cite{Mandel_Wolf_1995}. Each plane wave has a wavevector $\mathbf{k}=(k_x,k_y,k_z)^\mathrm{T}$ with magnitude $|\mathbf{k}|=\omega/c_0$, where $c_0$ is the speed of light in vacuum. The propagation direction of each plane wave is defined by the relative values of the components of $\mathbf{k}$. For convenience, we denote by $K$ the collection of transverse wavevectors $\mathbf{k}_\perp = (k_x, k_y)^\mathrm{T}$ satisfying $|\mathbf{k}_\perp| \leq k$. Given $\mathbf{k}_\perp$, we then have $|k_z| = \sqrt{k^2 - |\mathbf{k}_\perp|^2}$, which, notably, is real-valued for $\mathbf{k}_\perp \in K$.

For a point $\mathbf{r}$ in the far field of any sources or scatterers, evanescent components of the electric field are negligible. We may therefore write~\cite{Byrnes2021a}
\begin{align}\label{eq:angspec1}
    \mathbf{E}(\mathbf{r}) = \int_{K}[\mathbf{a}(\mathbf{k}_\perp)e^{i|k_z|z} + \mathbf{b}(\mathbf{k}_\perp)e^{-i|k_z| z}]\frac{e^{i\mathbf{k}_\perp\cdot\mathbf{r}_\perp}}{\sqrt{|k_z|}}\,\dee \mathbf{k}_\perp
\end{align}
for $\mathbf{r} \in \mathcal{R}^-$ where $\mathbf{r}_\perp = (x, y)^{\mathrm{T}}$. $\mathbf{E}(\mathbf{r})$ is defined similarly in $\mathbf{r} \in \mathcal{R}^+$ using vector functions $\mathbf{c}$ and $\mathbf{d}$ instead of $\mathbf{a}$ and $\mathbf{b}$ respectively. The vectors $\mathbf{a}, \mathbf{b}, \mathbf{c}$ and $\mathbf{d}$ define the polarization states of each plane wave.
 
To clarify the physical meaning of Eq.~(\ref{eq:angspec1}), consider the case in which sources are present in $\mathcal{R}^-$, but absent in $\mathcal{R}^+$. In this situation, no left-propagating (i.e., negative-$k_z$) plane waves exist in $\mathcal{R}^+$, and $\mathbf{d}$ therefore vanishes identically. The remaining coefficients can then be identified physically: $\mathbf{a}$ corresponds to the incident field, $\mathbf{b}$ to the field reflected by the scattering medium, and $\mathbf{c}$ to the transmitted field. More precisely, inspection of Eq.~(\ref{eq:angspec1}) shows that $\mathbf{a}(\mathbf{k}_\perp)/\sqrt{k_z}$ is proportional to the Fourier transform of the incident field evaluated in the plane $z=0$. This quantity should not be confused with the actual electric field at $z=0$, which lies inside the scattering medium and thus outside the domain of validity of Eq.~(\ref{eq:angspec1}). If sources are also present in $\mathcal{R}^+$, the coefficient $\mathbf{d}$ may be obtained analogously by considering the field generated by those sources in isolation.

Once $\mathbf{a}$ and $\mathbf{d}$ have been determined from the sources, $\mathbf{b}$ and $\mathbf{c}$ can be found using the continuous scattering matrix $\mathbf{S}(\mathbf{k}_\perp, \mathbf{k}'_\perp)$, which is a function of the incident $\mathbf{k}'_\perp$ and scattered $\mathbf{k}_\perp$ wavevectors. Defining the input and output vectors $\mathbf{I}(\mathbf{k}_\perp) = [\mathbf{a}(\mathbf{k}_\perp), \mathbf{d}(\mathbf{k}_\perp)]^\mathrm{T}$ and $\mathbf{O}(\mathbf{k}_\perp) = [\mathbf{b}(\mathbf{k}_\perp), \mathbf{c}(\mathbf{k}_\perp)]^\mathrm{T}$, $\mathbf{S}$ is defined through 
 \begin{align}
 \label{eq:ItoO}
     \mathbf{O}(\mathbf{k}_\perp) = \int_K \mathbf{S}(\mathbf{k}_\perp, \mathbf{k}'_\perp)\mathbf{I}(\mathbf{k}'_\perp)\,\dee\mathbf{k}'_\perp.
 \end{align}
Throughout this document, stacked vectors such as $\mathbf{I}(\mathbf{k}_\perp)$ and $\mathbf{O}(\mathbf{k}_\perp)$ are formed by concatenating their vector components into a single column vector; for example, if $\mathbf{a}$ and $\mathbf{d}$ each have three components, then $\mathbf{I}(\mathbf{k}_\perp)$ has six. It is convenient to write the scattering matrix in the block form
\begin{align}
\label{eq:Scontinuous}
    \mathbf{S}(\mathbf{k}_\perp, \mathbf{k}'_\perp) = \begin{pmatrix}
        \mathbf{r}(\mathbf{k}_\perp, \mathbf{k}'_\perp) &\mathbf{t}'(\mathbf{k}_\perp, \mathbf{k}'_\perp) \\
        \mathbf{t}(\mathbf{k}_\perp, \mathbf{k}'_\perp) &\mathbf{r}'(\mathbf{k}_\perp, \mathbf{k}'_\perp)
    \end{pmatrix},
\end{align}
where $\mathbf{r}$ and $\mathbf{r}'$ describe reflection, and $\mathbf{t}$ and $\mathbf{t}'$ describe transmission.

\subsection{Extended scattering channels and the discrete scattering matrix}
\label{sec:extended}

\begin{figure}[t] 
    \centering    
    \includegraphics[width=\columnwidth]{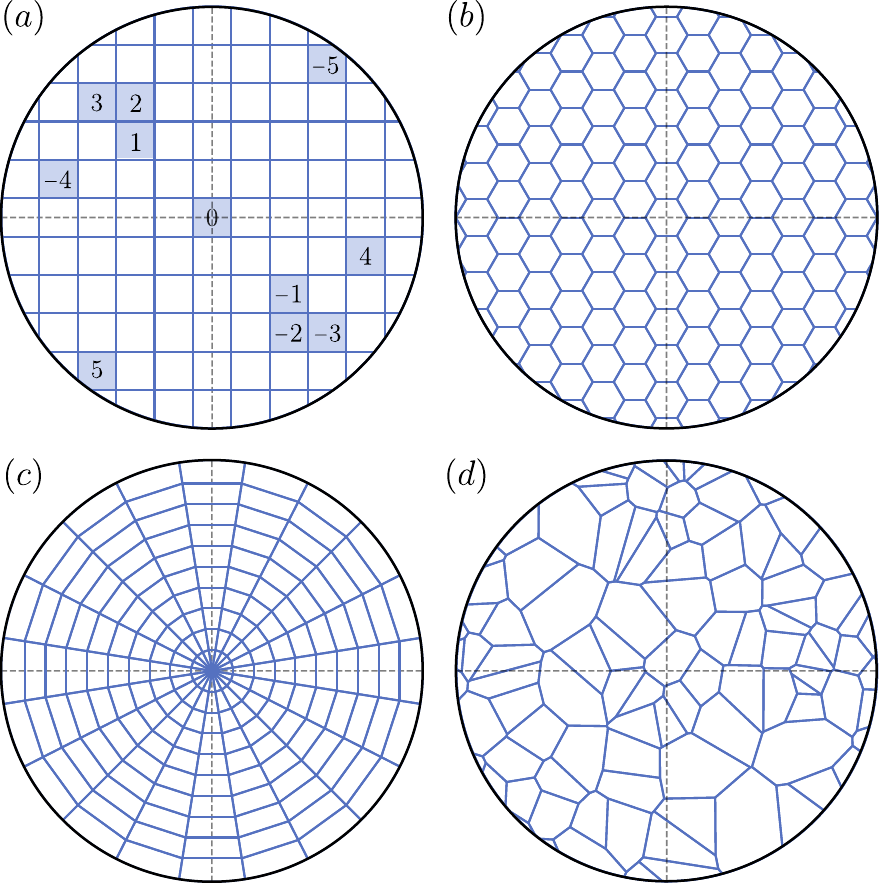} 
    \caption{Example scattering channel partitions of $K$: (a) square lattice, (b) hexagonal lattice, (c) polar grid, and (d) randomly generated Voronoi diagram. In (a), several regions are indexed according to the indexing scheme described in the text.\label{fig:partitions}}
    
\end{figure}

As is evident from Eq.~(\ref{eq:Scontinuous}), $\mathbf{S}$ is a function of the continuous variables $\mathbf{k}_\perp$ and $\mathbf{k}'_\perp$. To construct a finite-dimensional scattering matrix for numerical simulations, it is necessary to extract a finite collection of scattering channels from $K$. In Refs.~\cite{Byrnes14122022, ByrnesPhD}, this was achieved by overlaying a rectangular grid of wavevectors on $K$. Here, we instead consider a general partition $P = \{K_1, \dots, K_N\}$ of $K$ into $N$ disjoint regions $K_1, \dots, K_N$ where $\bigcup_{i=1}^N K_i = K$. Some example partitions with different region shapes are shown in Fig.~\ref{fig:partitions}. Physically, each region can be thought of as corresponding to a bundle of wavevectors, analogous to the bundle collected by an optical device whose photosensitive elements cover a non-zero angular range. Since each region extends over a range of wavevectors, we refer to them as \emph{extended scattering channels}. We note that while the majority of optical devices with discrete control elements, such as cameras, SLMs, DMDs, and deformable mirrors use rectangular arrays, some specialized devices adopt alternative geometries, such as hexagonal arrays~\cite{Ahmed2017, Zheng:23}, motivating the consideration of more general partitions.

Although any partition is possible in principle, we make two assumptions that substantially simplify the ensuing calculations. First, we assume that each region is a convex shape. Second, we assume that $P$ is invariant under inversion about the origin; that is, for each region $K_i$, the inverse region $-K_i = \{ -\mathbf{k}_\perp \mid \mathbf{k}_\perp \in K_i \}$ is also included in the partition. Note that since the regions are assumed to be disjoint, if the region $K_0$ contains the zero vector in its interior, then it must be its own inverse, i.e., $K_0 = -K_0$. Note that all regions in the four partitions shown in Fig.~\ref{fig:partitions} are convex. Moreover, the partitions in  Figs.~\ref{fig:partitions}(a)-(c) are invariant under inversion, whereas (d) is not. Note also that the central regions in Figs.~\ref{fig:partitions}(a) and (b) contain the origin and are themselves invariant under inversion, whereas Fig.~\ref{fig:partitions}(c) does not have a central region.

Inversion symmetry as defined above allows for a convenient indexing scheme for the regions. In particular, it is always possible to find subsets of either $n = (N-1)/2$ or $n = N/2$ regions $K_1, \dots, K_n$ where no two regions are inverse partners and no region contains the origin within its interior. A complete partition can then be formed by taking any such subset together with each member's inverse partner $K_{-i} := -K_i$ and, if present, the region $K_0$ containing the origin. The regions can be ordered as $K_{-n}, \dots, K_{-1}, K_0, K_1, \dots, K_n$, (omitting $K_0$ if no central region exists). Fig.~\ref{fig:partitions}(a) shows a partial numbering of the regions under this scheme. The central region is labeled $0$, a selected subset is labeled $1$ through $5$, and their inversion partners are labeled $-1$ through $-5$. It is clear that this numbering can be extended to cover the entire partition.

With a partition at hand, the integrals in Eqs.~(\ref{eq:angspec1}) can be separated into integrals over each region. In addition, using Eqs.~(\ref{eq:ItoO}) and (\ref{eq:Scontinuous}), the electric fields in $\mathcal{R}^-$ and $\mathcal{R}^+$ can be written purely in terms of the incident fields $\mathbf{a}$ and $\mathbf{d}$, and the different blocks of the scattering matrix. If, for example, sources only exist in $\mathcal{R}^-$, then for $\mathbf{r} \in \mathcal{R}^+$ we have
\begin{align}\label{eq:transcont}
    \mathbf{E}(\mathbf{r}) = \sum_{i,j}\int_{K_i \times K_j}\mathbf{t}(\mathbf{k}_{j\perp}, \mathbf{k}_{i\perp})\mathbf{a}(\mathbf{k}_{i\perp}) \frac{e^{i\mathbf{k}_j\cdot\mathbf{r}}}{\sqrt{|k_{jz}|}}\,\dee \mathbf{k}_{i\perp} \dee\mathbf{k}_{j\perp},
\end{align}
where $\mathbf{k}_{i\perp} \in K_i$, $\mathbf{k}_{j\perp} \in K_j$, and $K_i \times K_j$ is the Cartesian product of $K_i$ and $K_j$. We now assume that within each region (or pair of regions), $\mathbf{a}$ and $\mathbf{t}$ can be well approximated by their mean values so that Eq.~(\ref{eq:transcont}) can be written in the discrete form
\begin{align}
\label{eq:discrete}
    \mathbf{E}(\mathbf{r}) \approx \sum_{i,j}\mathbf{t}_{(j,i)}\mathbf{a}_i\phi_j(\mathbf{r}),
\end{align}
where
\begin{align}
    \mathbf{a}_i &= \frac{1}{w_i} \int_{K_i}\mathbf{a}(\mathbf{k}_{i\perp})\,\dee\mathbf{k}_{i\perp},\label{eq:discretea}\\
    \mathbf{t}_{(j,i)} &= \frac{1}{w_j} \int_{K_i \times K_j}\mathbf{t}(\mathbf{k}_{j\perp},\mathbf{k}_{i\perp})\,\dee\mathbf{k}_{i\perp}\dee\mathbf{k}_{j\perp},\label{eq:discrete-t}\\
        \phi_j(\mathbf{r}) &= \int_{K_j} \frac{1}{\sqrt{k_{jz}}}e^{ik_{jz} z}e^{i\mathbf{k}_{j\perp}\cdot\mathbf{r}_\perp}\,\dee\mathbf{k}_{j\perp},\label{eq:planewavebundles}
\end{align}
and $w_i = \int_{K_i}\dee\mathbf{k}_{i\perp}$ is the area of $K_i$. More generally, discrete versions of all of the blocks of the scattering matrix outlined in Eq.~(\ref{eq:Scontinuous}), namely $\mathbf{t}, \mathbf{t}', \mathbf{r}$ and $\mathbf{r}'$, can all be written in a form analogous to Eq.~(\ref{eq:discrete-t}).

To simplify the constraint imposed by energy conservation on the scattering matrix (see Appendix~\ref{sec:energy} for more information), we define a normalized scattering matrix $\widetilde{\mathbf{s}}_{(j,i)} = \sqrt{w_j / w_i}\mathbf{s}_{(j,i)}$, where $\mathbf{s}$ denotes $\mathbf{t}, \mathbf{t}', \mathbf{r}$, or $\mathbf{r}'$. Defined as such, $\widetilde{\mathbf{s}}_{(j,i)}$ is then given by
\begin{align}
\label{eq:sdisc}
    \widetilde{\mathbf{s}}_{(j,i)} = \frac{1}{\sqrt{w_iw_j}}\int_{K_i \times K_j}\mathbf{s}(\mathbf{k}_{j\perp},\mathbf{k}_{i\perp})\,\dee\mathbf{k}_{i\perp}\dee\mathbf{k}_{j\perp}.
\end{align}

We now turn to the size and structure of the discrete scattering matrix. Eq.~(\ref{eq:discretea}) shows that $\mathbf{a}_i$, is obtained by averaging $\mathbf{a}(\mathbf{k}_\perp)$ over the region $K_i$. For each wavevector, Gauss's law implies that $\mathbf{a}(\mathbf{k}_\perp)$ is orthogonal to $\mathbf{k}$, so it can be fully described by two independent components (commonly chosen as $s$ and $p$, or equivalently $\theta$ and $\phi$ polarizations)~\cite{mi06010m}. This allows $\mathbf{a}(\mathbf{k}_\perp)$ to be interpreted as a $2 \times 1$ Jones vector and $\mathbf{s}(\mathbf{k}_\perp, \mathbf{k}'_\perp)$ as a $2\times 2$ generalized Jones matrix~\cite{BYRNES2022127462}. The region-averaged vector $\mathbf{a}_i$ however, which is not associated with a single wavevector, does not satisfy this orthogonality property, and all three of its components must be retained. Consequently, $\mathbf{s}_{(j,i)}$ must be treated as $3 \times 3$ matrix. While there is no fundamental obstacle to tracking all three field components, the theory is simplified and the computational cost reduced if only two components are retained. With this in mind, we introduce an additional approximation by defining $\mathbf{a}_i$ through averaging only the field components, while keeping the coordinate vectors fixed; viz.,
\begin{align}
\label{eq:2dvs3d}
    \mathbf{a}_i \approx \frac{1}{w_i} \int_{K_i}[a_\theta(\mathbf{k}_{i\perp})\widehat{\mathbf{e}}_\theta(\mathbf{k}_{{i\perp},0})+ a_\phi(\mathbf{k}_{i\perp})\widehat{\mathbf{e}}_\phi(\mathbf{k}_{{i\perp},0})]\,\dee\mathbf{k}_{i\perp}.
\end{align}
Here, $\mathbf{k}_{i\perp,0} \in K_i$ is a fixed, arbitrary wavevector (for example, the centroid of $K_i$), and $\widehat{\mathbf{e}}_\theta(\mathbf{k}_{i\perp,0})$ and $\widehat{\mathbf{e}}_\phi(\mathbf{k}_{i\perp,0})$ are the corresponding unit polarization vectors. The quantities $a_\theta(\mathbf{k}_{i\perp}) = \mathbf{a}(\mathbf{k}_{i\perp}) \cdot \widehat{\mathbf{e}}_\theta(\mathbf{k}_{i\perp})$ and $a_\phi(\mathbf{k}_{i\perp}) = \mathbf{a}(\mathbf{k}_{i\perp}) \cdot \widehat{\mathbf{e}}_\phi(\mathbf{k}_{i\perp})$ denote the $\theta$ and $\phi$ components of $\mathbf{a}(\mathbf{k}_{i\perp})$, respectively. Extending this construction to $\mathbf{s}_{(j,i)}$ yields a $2\times 2$ matrix approximation obtained by averaging only the matrix elements corresponding to scattering between $\theta$ and $\phi$ polarization components.

Naturally, the approximations implicit in Eqs.~(\ref{eq:discrete}) and (\ref{eq:2dvs3d}) introduce errors whose magnitudes depend on the specific forms of the incident fields and scattering operators, as well as on the choice of partition. These errors decrease, however, as the partition is refined, i.e., as the number of regions increases and their individual sizes decrease.

Proceeding with the discussed approximations, we define the discrete scattering matrix $\mathbf{S}$ to be the $4N \times 4N$ matrix with block structure
\begin{align}
    \mathbf{S} &= \begin{pmatrix}
        \mathbf{r} &\mathbf{t}' \\
        \mathbf{t} &\mathbf{r}'
    \end{pmatrix},
\end{align}
where each block $\mathbf{s}$ (again denoting $\mathbf{t}, \mathbf{t}', \mathbf{r}$, or $\mathbf{r}'$) has the sub-block structure
\begin{align}
    \mathbf{s} &= 
\begin{pmatrix}
    \mathbf{s}_{(-n,-n)} & \cdots & \mathbf{s}_{(-n,n)} \\
    \vdots & \ddots & \vdots \\
    \mathbf{s}_{(n,-n)} & \cdots &   \begin{bmatrix}
        s_{(n,n)\theta\theta} &
s_{(n,n)\theta\phi}\\
s_{(n,n)\phi\theta} &s_{(n,n)\phi\phi}
\end{bmatrix} 
\end{pmatrix}.\label{eq:block-structure}
\end{align}
The bottom right sub-block $\mathbf{s}_{(n,n)}$ in Eq.~(\ref{eq:block-structure}) has been written explicitly to illustrate its internal $2\times 2$ polarization structure. A normalized scattering matrix $\widetilde{\mathbf{S}}$ can also be formed by replacing all instances of $\mathbf{s}$ with $\widetilde{\mathbf{s}}$. The discrete scattering matrix defined above allows us to write a simple matrix analog of Eq.~(\ref{eq:Scontinuous}). Defining the $4N$ component vectors
\begin{align}
    \mathbf{I} &= (\mathbf{a}_{-n},\dots,\mathbf
    a_n,\mathbf{d}_{-n},\dots,\mathbf{d}_n)^\mathrm{T},\label{eq:discI}\\
    \mathbf{O} &= (\mathbf{b}_{-n},\dots,\mathbf
    b_n,\mathbf{c}_{-n},\dots,\mathbf{c}_n)^\mathrm{T},\label{eq:discO}    
\end{align}
it is straightforward to see that $\mathbf{O} = \mathbf{S}\mathbf{I}$, which neatly links the averaged incident and scattered field components. 

\subsection{Physical and statistical model for the scattering medium}
\label{sec:statsmodel}
Having established the structure of the discrete scattering matrix, we now develop a physical and statistical model for the scattering medium, from which the statistical properties of the scattering matrix will be derived.

The scattering of a plane wave by a single scatterer can be described using the so-called amplitude matrix $\mathbf{A}(\mathbf{k}, \mathbf{k}'; \boldsymbol{\alpha})$, where $\mathbf{k}'$ is the incident plane wave's wavevector, $\mathbf{k}$ is the scattered wavevector, corresponding to a far field measurement direction, and $\boldsymbol{\alpha}$ is a vector of relevant physical parameters that determine the scatterer's properties \cite{Byrnes2021a}. For an isotropic sphere, for example, $\boldsymbol{\alpha}$ might consist of the size parameter and relative refractive index, as are usually defined in Mie theory \cite{Bohren1998}. For a non-spherical scatterer, $\boldsymbol{\alpha}$ will contain an alternate set of parameters that are sufficient to characterize its shape, orientation and any other relevant properties. Assuming that one can find $\mathbf{A}$ for each type of scatterer in the medium, it is possible to write down a simple expression for $\mathbf{s}(\mathbf{k}_\perp, \mathbf{k}'_\perp)$ in the single scattering regime, i.e., for $L$ smaller than the mean free path. We assume momentarily that the slab has finite transverse extent and contains $M$ scatterers in a finite volume $V$. If the $p$`th scatterer is located at position $\mathbf{r}_p$ and has physical parameters $\boldsymbol{\alpha}_p$, then it can be shown that \cite{Byrnes14122022}
\begin{align}
\label{eq:s-ss}
\mathbf{s}(\mathbf{k}_\perp, \mathbf{k}'_\perp) = \frac{1}{2\pi\sqrt{|k_{z}||k'_{z}|}}\sum_{p=1}^M\mathbf{A}^s_{p}(\mathbf{k}_\perp, \mathbf{k}'_\perp; \boldsymbol{\alpha}_p)e^{i\mathbf{r}_p\cdot(\mathbf{k}' - \mathbf{k})}.
\end{align}
Here, $\mathbf{A}_p^s$ denotes the amplitude matrix for the $p$th scatterer associated with the corresponding block $\mathbf{s}$ of the scattering matrix. Importantly, $\mathbf{A}_p^s$ is a function of \emph{transverse} wavevectors and is defined by evaluating the full amplitude matrix $\mathbf{A}_p(\mathbf{k}, \mathbf{k}'; \boldsymbol{\alpha}_p)$ with the appropriate signs taken for the $z$ components of the incident and scattered wavevectors. Specifically, suppressing the $\boldsymbol{\alpha}$ dependence for clarity, $\mathbf{A}_p^t(\mathbf{k}_\perp, \mathbf{k}'_\perp)
  = \mathbf{A}_p(\mathbf{k}_\perp, |k_z|, \mathbf{k}'_\perp, |k'_z|)$, $\mathbf{A}_p^{t'}(\mathbf{k}_\perp, \mathbf{k}'_\perp)
  = \mathbf{A}_p(\mathbf{k}_\perp, -|k_z|, \mathbf{k}'_\perp, -|k'_z|)$, $\mathbf{A}_p^r(\mathbf{k}_\perp, \mathbf{k}'_\perp)
  = \mathbf{A}_p(\mathbf{k}_\perp, -|k_z|, \mathbf{k}'_\perp, |k'_z|)$ and $\mathbf{A}_p^{r'}(\mathbf{k}_\perp, \mathbf{k}'_\perp)
  = \mathbf{A}_p(\mathbf{k}_\perp, |k_z|, \mathbf{k}'_\perp, -|k'_z|)$.

To model an ensemble of scattering matrices corresponding to different microscopic realizations of the scattering medium, it is necessary to specify the statistical properties of $\mathbf{s}(\mathbf{k}_\perp, \mathbf{k}'_\perp)$. The only sources of randomness are the positions of the scatterers $\mathbf{r}_p$, which influence the exponential factor in Eq.~(\ref{eq:s-ss}), and variations in the physical parameters associated with the scatterers $\boldsymbol{\alpha}_p$, which influence the form of $\mathbf{A}^s_p$. We suppose that all of these variables follow a probability density function $p(\mathbf{r}_1, \boldsymbol{\alpha}_1 ; \dots ; \mathbf{r}_M, \boldsymbol{\alpha}_M)$.

In the limit $M \to \infty$, the sum in Eq.~(\ref{eq:s-ss}) converges to a complex Gaussian random variable \cite{goodman2007speckle}. The statistics of the scattering matrix can therefore be fully described by the mean, covariance, and pseudo-covariance \cite{Schreier2010}, which are determined by the single and two-particle marginal density functions. To proceed, we make several statistical assumptions. First, we assume that the positions of the particles are independent of their physical parameters, implying that $p$ is separable according to
\begin{align}
\begin{split}
    p(\mathbf{r}_1, \boldsymbol{\alpha}_1; \dots ; \mathbf{r}_M, \boldsymbol{\alpha}_M)
    &= p_{\mathbf{r}_1,\dots,\mathbf{r}_M}(\mathbf{r}_1,\dots,\mathbf{r}_M) \\
    &\quad \times p_{\boldsymbol{\alpha}_1,\dots,\boldsymbol{\alpha}_M}
       (\boldsymbol{\alpha}_1,\dots,\boldsymbol{\alpha}_M).
\end{split}
\end{align}
This assumption implies that scatterers of all types are well-mixed within the medium. Second, we assume that the particle positions are independent and identically distributed (iid) random variables with uniform densities over the extent of the scattering medium, implying
\begin{align}
\label{eq:posdensity}
    p_{\mathbf{r}_1,\dots,\mathbf{r}_M}(\mathbf{r}_1,\dots,\mathbf{r}_M) = 1/V^M.
\end{align}
This assumption is clearly non-physical in reality as it allows for the possibility of multiple scatterers overlapping in space. It is a reasonable approximation, however, when the particles are distributed sparsely within the medium \cite{Tsang2000}. Under this assumption, the marginal single and two-particle joint densities are given simply by $p_\mathbf{r}(\mathbf{r}) = 1/V$ and $p_{\mathbf{r}_1,\mathbf{r}_2}(\mathbf{r}_1, \mathbf{r}_2) = 1/V^2$. Finally, we assume that the vectors $\boldsymbol{\alpha}_1, \dots \boldsymbol{\alpha}_M$ are also iid random variables, implying that their joint density is given by 
\begin{align}
    p_{\boldsymbol{\alpha}_1,\dots,\boldsymbol{\alpha}_M}(\boldsymbol{\alpha}_1,\dots,\boldsymbol{\alpha}_M) = \prod_{i=1}^M p_{\boldsymbol{\alpha}}(\boldsymbol{\alpha}_i),
\end{align}
where $p_{\boldsymbol{\alpha}}$ is the single particle marginal density. The two-particle marginal density is given by the product $p_{\boldsymbol{\alpha}_1, \boldsymbol{\alpha}_2}(\boldsymbol{\alpha}_1, \boldsymbol{\alpha}_2) = p_{\boldsymbol{\alpha}}(\boldsymbol{\alpha}_1)p_{\boldsymbol{\alpha}}(\boldsymbol{\alpha}_2)$. We note that this final assumption does not imply that the scatterers are monodisperse and only of a single kind. An equal mixture of scatterers with discrete properties $\boldsymbol{\alpha}_a$ and $\boldsymbol{\alpha}_b$, for example, could be represented by the density $p_{\boldsymbol{\alpha}}(\boldsymbol{\alpha}) = [\delta(\boldsymbol{\alpha} - \boldsymbol{\alpha}_a) + \delta(\boldsymbol{\alpha} - \boldsymbol{\alpha}_b)] /2$. We ultimately find that, within our model, the only remaining free parameter is $p_{\boldsymbol{\alpha}}(\boldsymbol{\alpha})$, and that all relevant statistics are fully determined once a choice for this distribution is specified.

\subsection{Mean normalized scattering matrix}
\label{sec:mean}
We now suppose that we are given a choice for $p_{\boldsymbol{\alpha}}(\boldsymbol{\alpha})$ and proceed to compute the statistics of the normalized scattering matrix.

Using Eq.~(\ref{eq:sdisc}), the mean of the sub-block $\widetilde{\mathbf{s}}_{(j,i)}$, which we denote by $\langle \widetilde{\mathbf{s}}_{(j,i)}\rangle$, is given by
\begin{align}
\label{eq:av-disc}
    \langle \widetilde{\mathbf{s}}_{(j,i)}\rangle = \frac{1}{\sqrt{w_iw_j}}\int_{K_i \times K_j}\langle\mathbf{s}(\mathbf{k}_{j\perp},\mathbf{k}_{i\perp})\rangle\,\dee\mathbf{k}_{i\perp}\dee\mathbf{k}_{j\perp},
\end{align}
where $\langle\mathbf{s}(\mathbf{k}_{j\perp},\mathbf{k}_{i\perp})\rangle$ is to be computed by averaging Eq.~(\ref{eq:s-ss}). Averaging over scatterer positions can be done explicitly. In the limit $V \to \infty$, we find that
\begin{align}\label{eq:t-cont-av}
\begin{split}
    \langle \mathbf{s}(\mathbf{k}_{j\perp}, \mathbf{k}_{i\perp}) \rangle = \,\,&\delta(\mathbf{k}_{i\perp} - \mathbf{k}_{j\perp})\frac{2\pi n_d L}{\sqrt{|k_{iz} ||k_{jz}|}}\\
    &\times\,\langle \mathbf{A}^s(\mathbf{k}_{j\perp}, \mathbf{k}_{i\perp})\rangle_{\boldsymbol{\alpha}}\mathrm{sinc}\big([\mathbf{k}_{ijz}\cdot \boldsymbol{\mu}^s]L/2\big),
\end{split}
\end{align}
where $\delta(\mathbf{k}_{i\perp} - \mathbf{k}_{j\perp}) = \delta(k_{ix} - k_{jx})\delta(k_{iy} - k_{jy})$, $\mathbf{k}_{ijz} = (|k_{iz}|, |k_{jz}|)^\mathrm{T}$, and $\boldsymbol{\mu}^s$ is a block-dependent constant vector that distributes positive and negative signs to the components of $\mathbf{k}_{ijz}$. In particular, the superscript $s$ matches the choice of block $\mathbf{s}$ and $\boldsymbol{\mu}^t = (1,-1)^\mathrm{T}$, $\boldsymbol{\mu}^{t'} = (-1,1)^\mathrm{T}$, $\boldsymbol{\mu}^r = (1,1)^\mathrm{T}$, and $\boldsymbol{\mu}^{r'} = (-1,-1)^\mathrm{T}$. In addition, 
\begin{align}
    \langle \mathbf{A}^s(\mathbf{k}_{j\perp}, \mathbf{k}_{i\perp})\rangle_{\boldsymbol{\alpha}} = \int \mathbf{A}^s(\mathbf{k}_{j\perp}, \mathbf{k}_{i\perp}; \boldsymbol{\alpha})p_{\boldsymbol{\alpha}}(\boldsymbol{\alpha})\,\dee\boldsymbol{\alpha}
\end{align}
and $n_d = M/V$ is the scatterer number density, which is assumed to remain finite as $V \to \infty$. Using Eqs.~(\ref{eq:av-disc}) and (\ref{eq:t-cont-av}), it is straightforward to calculate $\langle\widetilde{\mathbf{s}}_{(j,i)}\rangle$. Notably, since $K_i$ and $K_j$ are disjoint, the delta function is only non-vanishing when $i=j$. We find
\begin{align}\label{eq:t-av}
\begin{split}
   \langle \widetilde{\mathbf{s}}_{(j,i)}\rangle = \delta_{ij}\frac{2\pi n_d L}{w_i}\int_{K_i}&\langle \mathbf{A}^s(\mathbf{k}_{i\perp}, \mathbf{k}_{i\perp})\rangle_{\boldsymbol{\alpha}}\\
   &\times \frac{\mathrm{sinc}([\mathbf{k}_{iiz}\cdot \boldsymbol{\mu}^s]L/2)}{|k_{iz}|}\dee\mathbf{k}_{i\perp},
\end{split}
\end{align}
where $\delta_{ij}$ is the Kronecker delta. Note that the argument of the sinc in Eq.~(\ref{eq:t-av}) vanishes when $s \in \{t, t'\}$, simplifying the integrand. Recalling the form of the scattering matrix blocks in Eq.~(\ref{eq:block-structure}), Eq.~(\ref{eq:t-av}) shows that only diagonal sub-blocks of the blocks of the mean scattering matrix $\langle \mathbf{S}\rangle$ are non-zero. In the transmission blocks, these sub-blocks correspond to direct transmission, while, in the reflection blocks, these sub-blocks correspond to backscattering in the specular reflection direction.

\subsection{Normalized scattering matrix covariance and pseudo-covariance}
\label{sec:corr}
We now turn to the calculation of correlations between elements of the normalized scattering matrix. In order to simplify notation, we define an operator $C$ that takes a pair of matrices $\mathbf{s}_1$ and $\mathbf{s}_2$ and returns a matrix according to $C[\mathbf{s}_1, \mathbf{s}_2] = \mathrm{vec}(\mathbf{s}_1)\mathrm{vec}(\mathbf{s}_2)^\mathrm{T}$, where $\mathrm{vec}(\mathbf{s})$ is the vector formed by vectorizing $\mathbf{s}$ in row-major order. Furthermore, we define the covariance and pseudo-covariance
\begin{align}
    \mathrm{cov}[\mathbf{s}_1,\mathbf{s}_2] = \langle C[\mathbf{s}_1,\mathbf{s}^*_2] \rangle - C[\langle\mathbf{s}_1\rangle,\langle\mathbf{s}^*_2\rangle],\\
    \mathrm{pcov}[\mathbf{s}_1,\mathbf{s}_2] = \langle C[\mathbf{s}_1,\mathbf{s}_2] \rangle - C[\langle\mathbf{s}_1\rangle,\langle\mathbf{s}_2\rangle].
\end{align}
Ultimately, we require $\mathrm{cov}[\widetilde{\mathbf{s}}_{1(j,i)}, \widetilde{\mathbf{s}}_{2(v,u)}]$ and $\mathrm{pcov}[\widetilde{\mathbf{s}}_{1(j,i)}, \widetilde{\mathbf{s}}_{2(v,u)}]$
for all choices of channels $i,j,u$ and $v$, and all choices of scattering matrix blocks $\mathbf{s}_1$ and $\mathbf{s}_2$. Since the process for calculating the covariance and pseudo-covariance are essentially the same, we shall present the theory for the covariance and comment on pseudo-covariance briefly at the end.

Using again Eq.~(\ref{eq:sdisc}), we have
\begin{widetext}
\begin{align}\label{eq:long1}
\begin{split}
   \mathrm{cov}[\widetilde{\mathbf{s}}_{1(j,i)}, \widetilde{\mathbf{s}}_{2(v,u)}] =\frac{1}{\sqrt{w_iw_jw_uw_v}}\int_{K_i \times K_j\times K_u\times K_v}\mathrm{cov}[\mathbf{s}_1(\mathbf{k}_{j\perp},\mathbf{k}_{i\perp}), \mathbf{s}_2(\mathbf{k}_{v\perp},\mathbf{k}_{u\perp})]\,\dee\mathbf{k}_{i\perp}\dee\mathbf{k}_{j\perp}\dee\mathbf{k}_{u\perp}\dee\mathbf{k}_{v\perp}.
    \end{split}
\end{align}
\end{widetext}
The integrand of Eq.~(\ref{eq:long1}) can be found by averaging the product of two matrices given by Eq.~(\ref{eq:s-ss}). Since this product involves summation indices ranging over two different scatterers, the average must now be computed using the two particle marginal densities. Doing so, again in the limit $V \to \infty$, yields
\begin{widetext}
\begin{align}\label{eq:long2}
\begin{split}
&\mathrm{cov}[\mathbf{s}_1(\mathbf{k}_{j\perp}, \mathbf{k}_{i\perp}), \mathbf{s}_{2}(\mathbf{k}_{v\perp}, \mathbf{k}_{u\perp})]= \\
&\quad\frac{n_dL\delta(\mathbf{k}_{i\perp} - \mathbf{k}_{j\perp} - \mathbf{k}_{u\perp} + \mathbf{k}_{v\perp})}{\sqrt{|k_{iz}||k_{jz}||k_{uz}||k_{vz}|}} C[\langle \mathbf{A}^{s_1}(\mathbf{k}_{j\perp}, \mathbf{k}_{i\perp})\rangle_{\boldsymbol{\alpha}}, \langle\mathbf{A}^{s_2*}(\mathbf{k}_{v\perp}, \mathbf{k}_{u\perp})\rangle_{\boldsymbol{\alpha}}]\mathrm{sinc}\big([\mathbf{k}_{ijz}\cdot\boldsymbol{\mu}^{s_1} - \mathbf{k}_{uvz}\cdot\boldsymbol{\mu}^{s_2}]L/2\big),
\end{split}
\end{align}
\end{widetext}
where $\delta(\mathbf{k}_{i\perp} - \mathbf{k}_{j\perp} - \mathbf{k}_{u\perp} + \mathbf{k}_{v\perp}) = \delta(k_{ix} - k_{jx} - k_{ux} + k_{vx})\delta(k_{iy} - k_{jy} - k_{uy} + k_{vy})$. This delta function encodes the well-known memory effect \cite{Liu:19}, giving non-zero correlations for particular combinations of wavevectors satisfying the shift invariance relation 
\begin{align}\label{eq:mem-eff}
\mathbf{k}_{i\perp}-\mathbf{k}_{u\perp} = \mathbf{k}_{j\perp}-\mathbf{k}_{v\perp}.
\end{align}

In accordance with Eq.~(\ref{eq:long1}), $\mathrm{cov}[\widetilde{\mathbf{s}}_{1(j,i)}, \widetilde{\mathbf{s}}_{2(v,u)}]$ can be calculated by integrating the right hand side of Eq.~(\ref{eq:long2}) over the eight-dimensional space $K_i\times K_j\times K_u\times K_v$. Integrating the memory effect delta function eliminates two integration variables, leaving a six-dimensional integration domain. This domain, however, is non-trivial to parametrize and is not simply the product of three of the four original regions. In general, high-dimensional integrals involving delta functions can be evaluated using the so-called simple layer integral \cite{hörmander2015analysis}
\begin{align}\label{eq:coarea}
    \int_\Omega f(\mathbf{r})\delta(g(\mathbf{r}))\,\dee\mathbf{r} = \int_{\Omega\cap g^{-1}(0)}\frac{f(\mathbf{r})}{|\nabla g|}\,\dee \sigma,
\end{align}
where $\Omega$ is an $n$-dimensional domain and $f$ and $g$ are arbitrary, suitably defined functions. Critically, the domain of integration on the right hand side of Eq.~(\ref{eq:coarea}) is the intersection of $\Omega$ and the $(n-1)$-dimensional surface $g^{-1}(0)$, defined as the locus of points $\mathbf{r}$ for which $g(\mathbf{r}) =0$. The measure $\dee \sigma$ denotes the appropriate surface measure over this intersection. For our purposes, we use Eq.~(\ref{eq:coarea}) twice, once for each delta function.

Beginning with the domain $K_i \times K_j \times K_u \times K_v$, we note that the memory effect enforces two linear constraints, each of which can be represented by a hyperplane. We shall denote these hyperplanes by $\Pi_x$ and $\Pi_y$. The first, $\Pi_x$, corresponds to the $x$ component constraint $k_{ix} - k_{jx} - k_{ux} + k_{vx} = 0$, while $\Pi_y$ encodes the analogous constraint on the $y$ components. The resulting integration domain after integrating the delta functions is therefore the six-dimensional shape $(K_i \times K_j \times K_u \times K_v) \cap \Pi_x \cap \Pi_y$. It remains to be demonstrated how this space may be determined and parameterized for numerical integration. Here we adopt an algebraic technique based on the double description method for convex polytopes \cite{10.1007/3-540-61576-8_77}.

Recalling that each region is a convex set, it follows straightforwardly that the product $K_i \times K_j \times K_u \times K_v$ is also convex. Furthermore, since each hyperplane is convex, the resulting intersection is also convex. A list of vertices for  $K_i \times K_j \times K_u \times K_v$ can be obtained by enumerating all possible products of vertices of the four regions. $K_i \times K_j \times K_u \times K_v$ can then be identified as the unique convex hull of the resulting set of vertices. This set of vertices constitutes a V (vertex) representation of the convex polytope $K_i \times K_j \times K_u \times K_v$. Next, it is possible to convert to a H (halfspace) representation using, e.g., cddlib \cite{cddlib}. The H representation consists of a list of $n_f$ inequalities, each of the form $\mathbf{n}_i \cdot \mathbf{x} + \mathbf{d}_i \geq 0$, where $\mathbf{x} = (k_{ix}, k_{iy}, k_{jx}, k_{jy}, k_{ux}, k_{uy}, k_{vx}, k_{vy})^\mathrm{T}$ and each $\mathbf{n}_i$ and $\mathbf{d}_i$ are eight-component constant vectors. Each inequality of this form forces $\mathbf{x}$ to lie on one side of a hyperplane containing one of the seven-dimensional facets of $K_i \times K_j \times K_u \times K_v$. It is now straightforward to see that the H representation of the desired integration domain is given by the augmented set of constraints
\begin{align}\label{eq:cases1}
    \begin{cases}
        &\mathbf{n}_i \cdot \mathbf{x} + \mathbf{d}_i \geq 0\quad\quad\quad\quad\quad\quad\quad 1 \leq i \leq n_f,\\
        &k_{ix} - k_{jx} - k_{ux} + k_{vx} = 0\\
        &k_{iy} - k_{jy} - k_{uy} + k_{vy} = 0,
    \end{cases}
\end{align}
which imposes the memory effect conditions onto the original domain.
We then convert the augmented H representation in Eq.~(\ref{eq:cases1}) to a V representation, which yields a list of vertices of the desired intersection.

The vertices produced by the discussed algorithm outline a six-dimensional subspace embedded in an eight-dimensional ambient space. We denote a general point in the ambient space by $\mathbf{r}^8$ and a general point in the subspace by $\mathbf{r}^8_s$. Given the coordinates $k_{ix}, k_{iy}, k_{jx}, k_{jy}, k_{ux}, k_{uy}, k_{vx}$ and $k_{vy}$ for the ambient space, we choose the first six of these variables as intrinsic coordinates for the subspace and denote a general point in terms of these coordinates as $\mathbf{r}^6_s$. Explicitly,
\begin{align}
    \mathbf{r}^8 = \begin{pmatrix}
        k_{ix}\\
        k_{iy}\\
        k_{jx}\\
        k_{jy}\\
        k_{ux}\\
        k_{uy}\\
        k_{vx}\\
        k_{vy}\\
    \end{pmatrix},\mathbf{r}^8_s = \begin{pmatrix}
        k_{ix}\\
        k_{iy}\\
        k_{jx}\\
        k_{jy}\\
        k_{ux}\\
        k_{uy}\\
        -k_{ix}+k_{jx}+k_{ux}\\
        -k_{iy}+k_{jy}+k_{uy}\\
    \end{pmatrix},
\mathbf{r}^6_s = \begin{pmatrix}
        k_{ix}\\
        k_{iy}\\
        k_{jx}\\
        k_{jy}\\
        k_{ux}\\
        k_{uy}\\
    \end{pmatrix}.
\end{align}
The surface measure for the subspace is then given by $\dee \sigma = \sqrt{\det(\mathbf{G})}\prod_{i=1}^6\dee x_i$, where $x_i$ enumerates the coordinates in $\mathbf{r}^6_s$ and $\mathbf{G}$ is the Gram matrix with general element $G_{ij} = \frac{\partial \mathbf{r}^8_s}{\partial x_i} \cdot \frac{\partial \mathbf{r}^8_s}{\partial x_j}$ \cite{Courant1989}. In our particular case, $\sqrt{\det(\mathbf{G})} = 4$. The function $g$ in each application of Eq.~(\ref{eq:coarea}) is the delta function argument, for example $k_{ix} - k_{jx} - k_{ux} + k_{vx}$ in the case of $\Pi_x$. We therefore find that $|\nabla g| = 2$ for each use of Eq.~(\ref{eq:coarea}), resulting in a factor of four that cancels $\sqrt{\det(\mathbf{G})}$. This means that the integral in Eq.~(\ref{eq:long1}) can ultimately be evaluated by integrating Eq.~(\ref{eq:long2}) over the domain returned from the double description method with respect to $\mathbf{k}_{i\perp}, \mathbf{k}_{j\perp}$ and $\mathbf{k}_{u\perp}$, enforcing $\mathbf{k}_{v\perp} = -\mathbf{k}_{i\perp} + -\mathbf{k}_{j\perp} + -\mathbf{k}_{u\perp}$ in the integrand. In general, the integration domain will not be of a shape automatically amenable to common integration techniques. Since the domain is convex, however, it can be readily decomposed into simplices, and the integral can then be evaluated over each simplex using standard cubature schemes \cite{stroud1971approximate}.

\begin{figure}[t]
    \includegraphics[width=\columnwidth]{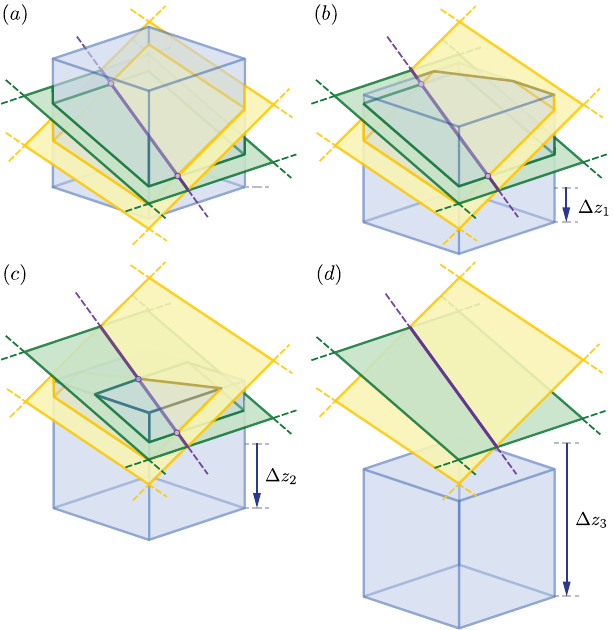} 
    \caption{3-dimensional analogue of the intersection $(K_i \times K_j \times K_u \times K_v)\cap \Pi_x \cap \Pi_y$. Going from (a) to (d), the line segment associated with the cube's vertical axis is translated vertically downwards, causing the cube to shift downwards relative to the planes and the intersection to decrease in length.}
    \label{fig:cube}
\end{figure}

To demonstrate some of the complexities and physical implications of the geometry described in this section, we consider a lower dimensional analog of our original problem, which is depicted in Fig.~\ref{fig:cube}. This figure shows various cases of the intersection of a cube with two planes. Here, the cube can be viewed as the Cartesian product of three orthogonal line segments, $L_1 \times L_2 \times L_3$, and serves as an analogy for the 8-dimensional space $K_i \times K_j \times K_u \times K_v$. Naturally, the two planes correspond to $\Pi_x$ and $\Pi_y$, and the intersection of these planes with the cube thus corresponds to $(K_i \times K_j \times K_u \times K_v)\cap \Pi_x \cap \Pi_y$. While the planes are fixed, the position of the cube relative to the planes is determined by the positions of the three line segments. Fig.~\ref{fig:cube}(a) shows a particular configuration of the line segments where the intersection extends from the cube's front face to its back face. Figs.~\ref{fig:cube}(b)-(d) show variations of this configuration where the line segment corresponding to the vertical axis of the cube is displaced downwards from its initial position by $\Delta z_1$, $\Delta z_2$, and $\Delta z_3$ respectively, which causes the cube to shift vertically downwards by the same amount. In our original problem, this shifting of a line segment corresponds to taking a particular quadruple of regions $K_i$,$K_j$, $K_u$ and $K_v$, and replacing one, e.g., $K_v$, with another region $K'_v$ slightly displaced from $K_v$. Importantly this displacement affects the length of the resulting intersection, which is representative of the size of the integration domain, and thus closely tied to the strength of $\mathrm{cov}[\widetilde{\mathbf{s}}_{1(j,i)}, \widetilde{\mathbf{s}}_{2(v,u)}]$. If the displacement is sufficiently small, as is the case in Fig.~\ref{fig:cube}(b), the intersection still extends between the front and back faces of the cube and is thus the same size as in the initial configuration. As the displacement increases, however, the rear end of the intersection eventually crosses a cube edge, as can be seen in going from Fig.~\ref{fig:cube}(b) to Fig.~\ref{fig:cube}(c). In Fig.~\ref{fig:cube}(c), the intersection endpoint that was initially on the cube's rear face is now on its top face and the intersection is thus shorter than before. If the displacement of the segment is made sufficiently large, as is the case in Fig.~\ref{fig:cube}(d), there is no intersection between the cube and the planes. In the original problem, this corresponds to a quadruple of regions for which no combination of wavevectors are able to satisfy the memory effect conditions.

Finding $\mathrm{pcov}[\widetilde{\mathbf{s}}_{1(j,i)}, \widetilde{\mathbf{s}}_{2(v,u)}]$ poses no additional conceptual difficulties beyond those already encountered. The procedure is essentially identical, with only minor modifications required in the expressions of Eq.~(\ref{eq:long2}). In particular, the argument of the delta function should be replaced with $\mathbf{k}_{i\perp}-\mathbf{k}_{j\perp} + \mathbf{k}_{u\perp}-\mathbf{k}_{v\perp}$, whose vanishing corresponds to the lesser-known conjugate memory effect \cite{Kawanishi:99}. Additionally, within the sinc function, one should replace $\mathbf{k}_{uvz} \to -\mathbf{k}_{uvz}$ and remove the complex conjugation on the latter amplitude matrix factor.

For a general partition, there is no simple relationship between the integration domains associated with covariance and pseudo-covariance calculations. For inversion-symmetric partitions, however, a straightforward connection exists. Suppose that in the calculation of $\mathrm{cov}[\widetilde{\mathbf{s}}_{1(j,i)}, \widetilde{\mathbf{s}}_{2(v,u)}]$, the domain $(K_i \times K_j \times K_u \times K_v)\cap \Pi_x \cap \Pi_y$ was obtained. Then, by comparing the wavevector conditions for the memory and conjugate memory effects, it is clear that the exact same integration domain would also be obtained in the calculation of $\mathrm{pcov}[\widetilde{\mathbf{s}}_{1(j,i)}, \widetilde{\mathbf{s}}_{2(-v,-u)}]$. Consequently, enumerating all integration domains for covariance calculations automatically yields the corresponding domains required for pseudo-covariance calculations.

\subsection{Random matrix generation}
\label{sec:generation}
Having discussed the statistics of the scattering matrix elements, we now outline the workflow for computing these statistics and generating random matrices. The general strategy is to exploit various symmetries to reduce the total number of required calculations.

For certain classes of scatterers, one can exploit symmetries associated with the scatterer’s physical properties to reduce computational cost. For example, it is well-known that the scattering amplitudes of an isotropic sphere depend only on the scattering angle, rather than on the specific choice of incident and scattered wavevectors \cite{van1981light}. Symmetries of this kind, however, are not guaranteed to extend to more general classes of scatterers and must be implemented on a case-by-case basis. A far more robust symmetry is reciprocity, which applies to the vast majority of optical systems of interest. It follows from reciprocity that each sub-block of $\mathbf{S}$ has a reciprocal partner that is related to it by a deterministic relationship, regardless of the nature of the scattering medium. Consequently, it is only ever necessary to compute one member of any reciprocal pair, which reduces the number of required statistical calculations by approximately a half. The details of this symmetry are provided in Sec.~\ref{sec:reciprocity}.

We start by identifying a minimal set of independent sub-blocks $\mathcal{B}$ of $\mathbf{S}$ on the basis of reciprocity. Since the reciprocal partner of any sub-block of $\mathbf{t}$ lies within $\mathbf{t}'$, we take the full set of sub-blocks of $\mathbf{t}$ and none from $\mathbf{t}'$. The reciprocal partner of a sub-block of $\mathbf{r}$ can be seen to be the sub-block obtained by reflecting the location of the original sub-block across the anti-diagonal of $\mathbf{r}$. Sub-blocks that lie on the anti-diagonal of $\mathbf{r}$ are their own reciprocal partners and correspond physically to direct backscattering (where one observes the coherent backscattering effect). We therefore take all sub-blocks lying on or above the anti-diagonal of $\mathbf{r}$, which corresponds to sub-blocks $\mathbf{r}_{(j,i)}$ such that $i+j \leq 0$. An analogous procedure applies to $\mathbf{r}'$. This gives a total of $|\mathcal{B}| = N^2 + 2[N(N+1)/2] = 2N^2 + N$ sub-blocks, which is slightly more than half of the $4N^2$ sub-blocks in $\mathbf{S}$.

Mean sub-blocks must be calculated for all transmission sub-blocks in $\mathcal{B}$ lying on the diagonal of $\mathbf{t}$. In $\mathbf{r}$ (and $\mathbf{r}'$), sub-blocks in $\mathcal{B}$ with non-zero mean extend along the diagonal from the top left corner to the center of the matrix where the diagonal meets the anti-diagonal. When the partition possesses a central region, there are therefore in total $N + 2(n+1) = 2N-1$ sub-blocks for which the mean must be calculated ($2N$ in the case that the partition does not possess a central region).

A priori, any member of $\mathcal{B}$ could covary with any other. Since the order in which the covariance of two blocks is calculated doesn't matter (owing to the Hermitian symmetry of the covariance matrix), the total number of potential covariance calculations is therefore $\binom{|\mathcal{B}|}{2} \sim 2N^4$. Once a particular covariance has been calculated, reciprocity also allows one to automatically obtain three additional covariances. Suppose, for example, $\mathrm{cov}[\widetilde{\mathbf{s}}_{1(j,i)}, \widetilde{\mathbf{s}}_{2(v,u)}]$ has been found. It is, again in light of Sec.~\ref{sec:reciprocity}, relatively straightforward to show that 
\begin{align}
\mathrm{cov}[\widetilde{\mathbf{s}}_{1(-i,-j)}, \widetilde{\mathbf{s}}_{2(v,u)}] &= \mathbf{R}\,\mathrm{cov}[\widetilde{\mathbf{s}}_{1(j,i)}, \widetilde{\mathbf{s}}_{2(v,u)}],\\
\mathrm{cov}[\widetilde{\mathbf{s}}_{1(j,i)}, \widetilde{\mathbf{s}}_{2(-u,-v)}] &= \mathrm{cov}[\widetilde{\mathbf{s}}_{1(j,i)}, \widetilde{\mathbf{s}}_{2(v,u)}]\mathbf{R},\\
\mathrm{cov}[\widetilde{\mathbf{s}}_{1(-i,-j)}, \widetilde{\mathbf{s}}_{2(-u,-v)}] &= \mathbf{R}\,\mathrm{cov}[\widetilde{\mathbf{s}}_{1(j,i)}, \widetilde{\mathbf{s}}_{2(v,u)}]\mathbf{R},
\end{align}
where
\begin{align}
    \mathbf{R} = \begin{pmatrix}
        1 &0 &0 &0\\
        0 &0 &-1 &0\\
        0 &-1 &0 &0\\
        0 &0 &0 &1
    \end{pmatrix}.
\end{align}
Analogous relationships hold for pseudo-covariances.

Although scattering matrix symmetries help reduce the total number of correlation calculations, it is still not immediately obvious which pairs of scattering matrix sub-blocks within $\mathcal{B}$ actually covary. As discussed in Sec.~\ref{sec:corr}, the existence (or lack thereof) of correlations is intricately tied to the geometry of the four regions involved in any given correlation. The vast majority of pairs of elements in $\mathcal{B}$, however, do not covary, and thus require no explicit calculation. Since the number of elements in $\mathcal{B}$ grows rapidly with scattering matrix size and accurate integration can be computationally expensive, it is therefore advantageous to first filter the set of potential correlations before attempting to evaluate all integrals. This filtering problem is difficult in general, but a natural approach is to compute the volume $\sigma$ of the integration domain and compare it against a threshold value $\sigma_0$, proceeding with the integration only if $\sigma > \sigma_0$. In practice, choosing a value of $\sigma_0$ involves a trade-off between computational speed and simulation accuracy; larger values of $\sigma_0$ result in fewer calculations, but retain only the strongest correlations. Precisely which values of $\sigma_0$ are appropriate for capturing the most relevant correlations depends intricately on the partition geometry, as different geometries can lead to widely varying distributions of $\sigma$. When $\mathcal{B}$ is very large, calculating $\sigma$ for each candidate covariance can also be computationally prohibitive. In such cases, it can be useful to approximate $\sigma$ using alternative methods. One example of such a method, appropriate for arbitrary partitions, is outlined in Appendix~\ref{sec:minkowski}.

\begin{figure}[t]
\includegraphics[width=\columnwidth]{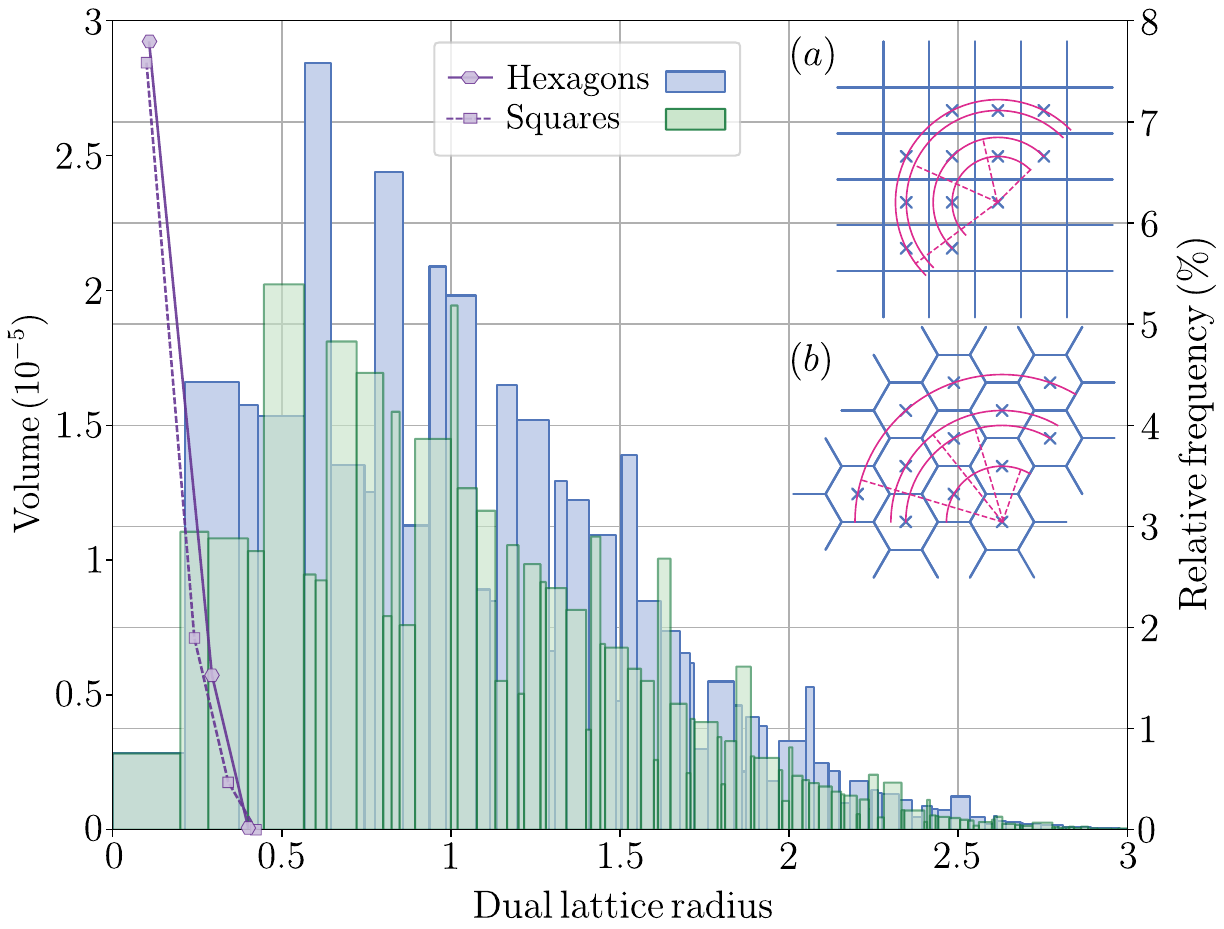} 
    \caption{Left $y$-axis: 6-dimensional intersection volume $\sigma$ as a function of the dual-lattice radius $|\mathbf{q}_{ijuv,0}|$.
Right $y$-axis: relative frequency of region quadruples at a given dual-lattice radius.
Insets: examples of dual-lattice radii for (a) square and (b) hexagonal grids.}
    \label{fig:lattice-bars}
\end{figure}

For regular lattice partitions, such as the rectangular and hexagonal lattices shown in Fig.~\ref{fig:partitions}(a) and (b), the situation simplifies significantly. Consider, for example, the interior of the square lattice in Fig.~\ref{fig:partitions}(a) temporarily ignoring truncated regions near the circular boundary, and its dual lattice, defined as the lattice of midpoints of each square region.
Clearly, for any quadruple of regions, the dual vector $\mathbf{q}_{ijuv,0} = \mathbf{k}_{i\perp,0} - \mathbf{k}_{j\perp,0} - \mathbf{k}_{u\perp,0} + \mathbf{k}_{v\perp,0}$, formed from the midpoints of each region, also lies on the dual lattice, albeit potentially extending beyond the circular boundary of the original grid. Importantly, the volume $\sigma$ is uniquely determined by the magnitude $|\mathbf{q}_{ijuv,0}|$. Since $\mathbf{q}_{ijuv,0}$ lies on the dual lattice, its magnitude can only take a relatively small number of distinct values, corresponding precisely to the radii of dual-lattice points. The inset Fig.~\ref{fig:lattice-bars}(a) shows the first few of these radii for the square lattice, while Fig.~\ref{fig:lattice-bars}(b) shows the corresponding radii for a hexagonal lattice. Naturally, as the radius increases, $\sigma$ decreases, quickly reaching $0$ after only a few radii, as shown by the purple lines in Fig.~\ref{fig:lattice-bars}, which were generated from square and hexagonal lattices each with approximately $100$ regions. In practice, this means that $\sigma$ for any particular quadruple of regions can be determined directly from the magnitude of the corresponding dual vector $\mathbf{q}_{ijuv,0}$, once the volume has been computed explicitly for one representative case. This allows for rapid filtering of candidate quadruples. Moreover, as illustrated by the bar chart in Fig.~\ref{fig:lattice-bars}, which show the relative frequencies of quadruples for each radius, the first few bars, corresponding to nonzero values of $\sigma$, represent only a small fraction of the total number of candidates. This filtering process thus drastically reduces the overall computational cost.

Once all covariances and pseudo-covariances have been calculated, the results can be assembled into the covariance and pseudo-covariance matrices, $\mathbf{C}$ and $\mathbf{P}$, each of size $16N^2 \times 16N^2$. These matrices can be viewed as $4 \times 4$ block matrices, where each block describes the correlations between a pair of blocks from $\mathbf{S}$. Each such block further contains $N^2$ sub-blocks of size $4 \times 4$, describing the correlations between corresponding $2\times2$ sub-blocks of $\mathbf{S}$. Random generation of scattering matrices can be achieved by constructing an augmented, real-valued covariance matrix with block form \cite{Schreier2010}
\begin{align}
    \boldsymbol{\Sigma} = \frac{1}{2}\begin{pmatrix}
        \mathrm{Re}(\mathbf{C} + \mathbf{P}) & \mathrm{Im}(-\mathbf{C} + \mathbf{P})\\[2mm]
        \mathrm{Im}(\mathbf{C} + \mathbf{P}) & \mathrm{Re}(\mathbf{C} - \mathbf{P})
    \end{pmatrix}.
\end{align}
Feeding $\boldsymbol{\Sigma}$ into a real-valued, zero-mean, Gaussian random number generator yields a vector of length $32N^2$, the first and second halves of which contain the real and imaginary parts of the elements of a random normalized scattering matrix. These elements can then be carefully rearranged and assembled into a matrix  $\Delta\widetilde{\mathbf{S}}$.
Finally, having also assembled the mean normalized scattering matrix $\langle \widetilde{\mathbf{S}} \rangle $ from the previously computed mean values, the mean-corrected scattering matrix is obtained as $\widetilde{\mathbf{S}} = \langle \widetilde{\mathbf{S}} \rangle + \Delta \widetilde{\mathbf{S}}$.

Energy conservation (see Appendix~\ref{sec:energy}) dictates that scattering matrices should be unitary. A randomly generated matrix from a Gaussian distribution, however, is unitary with probability zero. To enforce unitarity, we compute the singular value decomposition $\widetilde{\mathbf{S}} = \mathbf{U}\mathbf{D}\mathbf{V}^\dagger$, and take $\mathbf{U}\mathbf{V}^\dagger$ as our final scattering matrix. This procedure preserves the reciprocity built into $\widetilde{\mathbf{S}}$, but alters the matrix relative to the original randomly generated one. \red{Errors introduced by this process have been studied in earlier works~\cite{ByrnesPhD} and alternative strategies that circumvent this problem may be sought in the future.}

Random matrices generated using the above \red{(and summarized below)} describe scattering media with physical properties given by the choice of input parameters. Most notably, our model is based on single scattering theory, limiting the medium thickness $L$ to less than a mean free path. Scattering matrices for thicker media, however, can be obtained by cascading many instances of randomly generated thin media until the desired thickness is reached. This can be achieved through standard product formulae for scattering matrices, or using transfer matrices. Additional propagation matrices, which account for the phase accrued over a cascade, are also required; their construction has been described elsewhere~\cite{ByrnesPhD, Byrnes14122022} and shall not be reproduced here. We also employ a multi-pool, bootstrapping approach to random matrix sampling, which significantly reduces the number of required matrix products. This approach was originally introduced in Refs.~\cite{ByrnesPhD, Byrnes14122022} on an ad-hoc basis, but has since been analyzed in greater detail in a more recent statistical study~\cite{Byrnes2024a}.

\begin{redtext}
    
\subsection{Summary}\label{sec:summary}
Having described each component of the random matrix generation algorithm, we now summarize the workflow from setting up the matrix generator to  conducting statistical studies. The procedure outlined below is the one implemented in our code~\cite{rmtnew}.

Before performing scattering simulations, a random matrix generator must first be constructed. Although this step can be computationally expensive, particularly for fine grids, it is carried out only once for a fixed set of parameters (e.g., Fourier-space grid, scatterer type, etc.). The resulting generator can be stored and reused by loading it from disk. The construction procedure is as follows:
\begin{enumerate}
    \item Generate the Fourier-space partition.
    \item Identify the minimal set of scattering matrix indices for which statistical quantities must be computed.
    \item Compute the mean values of the scattering matrix elements and store the results on disk.
    \item Compute the covariance of the scattering matrix elements.
    \item Assemble the results into the real covariance matrix $\boldsymbol{\Sigma}$ and store it on disk.
\end{enumerate}
The computed mean vector and covariance matrix constitute the key ingredients of the random matrix generator. In practice, we additionally compute the Cholesky decomposition of $\boldsymbol{\Sigma}$ to facilitate the generation of correlated random variables~\cite{glasserman2004monte}. We  note that when a simple cubature scheme, such as the midpoint rule, is used, only a small set of $\mathbf{A}$ matrices and integration domain volumes are required to complete all calculations. These values can be calculated in advance of steps 3 and 4 and held in memory for repeated use.

Generating a single random matrix corresponding to a medium of thickness $L$ involves the following steps:
\begin{enumerate}
    \item Load the mean scattering matrix and the Cholesky decomposition of $\boldsymbol{\Sigma}$ from disk.
    \item Generate correlated Gaussian random variables from standard normal samples using the Cholesky factor.
    \item Assemble the generated values into a scattering matrix.
    \item Add the mean scattering matrix to the random realization.
    \item Symmetrize the resulting scattering matrix using the SVD method.
\end{enumerate}
In practice, many matrices can be generated vectorially in a single batch. This additionally reduces the I/O overhead associated with repeatedly loading the Cholesky decomposition from disk.

Finally, the matrix cascade procedure consists of the following steps:
\begin{enumerate}
    \item Generate a pool of single-layer scattering matrices using the random matrix generator.
    \item Generate a pool of multi-layer scattering matrices using a bootstrapping approach.
    \item Use the multi-layer pool to construct cascaded matrices for the desired medium thicknesses.
    \item Compute the desired statistical quantities from the cascaded scattering matrices.
\end{enumerate}

\end{redtext}

\section{Numerical examples}\label{sec:numerical}
In this section, we demonstrate some of the capabilities of our code through simulations of polarized waves interacting with random scattering media. For simplicity, the examples considered here involve random media composed of isotropic, spherical scatterers, for which the scattering properties are well known~\cite{Bohren1998}. 

\subsection{Fields in real and Fourier space}\label{sec:fields}
\label{sec:fieldsscat}
We first demonstrate the scattering of a vectorial beam by a random medium. In particular, we consider a Hermite--Gaussian beam of order $\mathrm{HG}_{11}$ incident upon a thin scattering layer composed of small, Rayleigh scatterers with size parameter $0.1$ and relative refractive index $1.2$. We choose a polar partition, similar to that shown in Fig.~\ref{fig:partitions}(c), to represent the fields. For our partition, we choose a radial spacing of $\Delta r = 0.05\,k$ and an angular spacing of $\Delta \theta = 2\pi/40$, resulting in a total of $800$ extended channels. \red{Only auto-correlations of scattering matrix sub-blocks were included at this stage, which means that memory effect correlations were `turned off'.} Assuming the incident beam is polarized along the $y$ direction, the Fourier-space representation of the beam in the plane of the waist can be written as~\cite{https://doi.org/10.48550/arxiv.physics/0410021}
\begin{align}
\label{eq:gaussian-beam}
    \mathbf{E}(k_x,k_y) = k_x k_y 
    e^{-\frac{w_0^2}{4}(k_x^2+k_y^2)}
    \left(\widehat{\mathbf{e}}_y - [\widehat{\mathbf{e}}_y\cdot\widehat{\mathbf{e}}_k]\widehat{\mathbf{e}}_k\right),
\end{align}
where $w_0$ is the beam waist radius, $\widehat{\mathbf{e}}_y$ is a unit vector pointing in the $y$-direction, and $\widehat{\mathbf{e}}_k$ is a unit vector parallel to $\mathbf{k}$. 

\begin{figure}[t]
\includegraphics[width=\columnwidth]{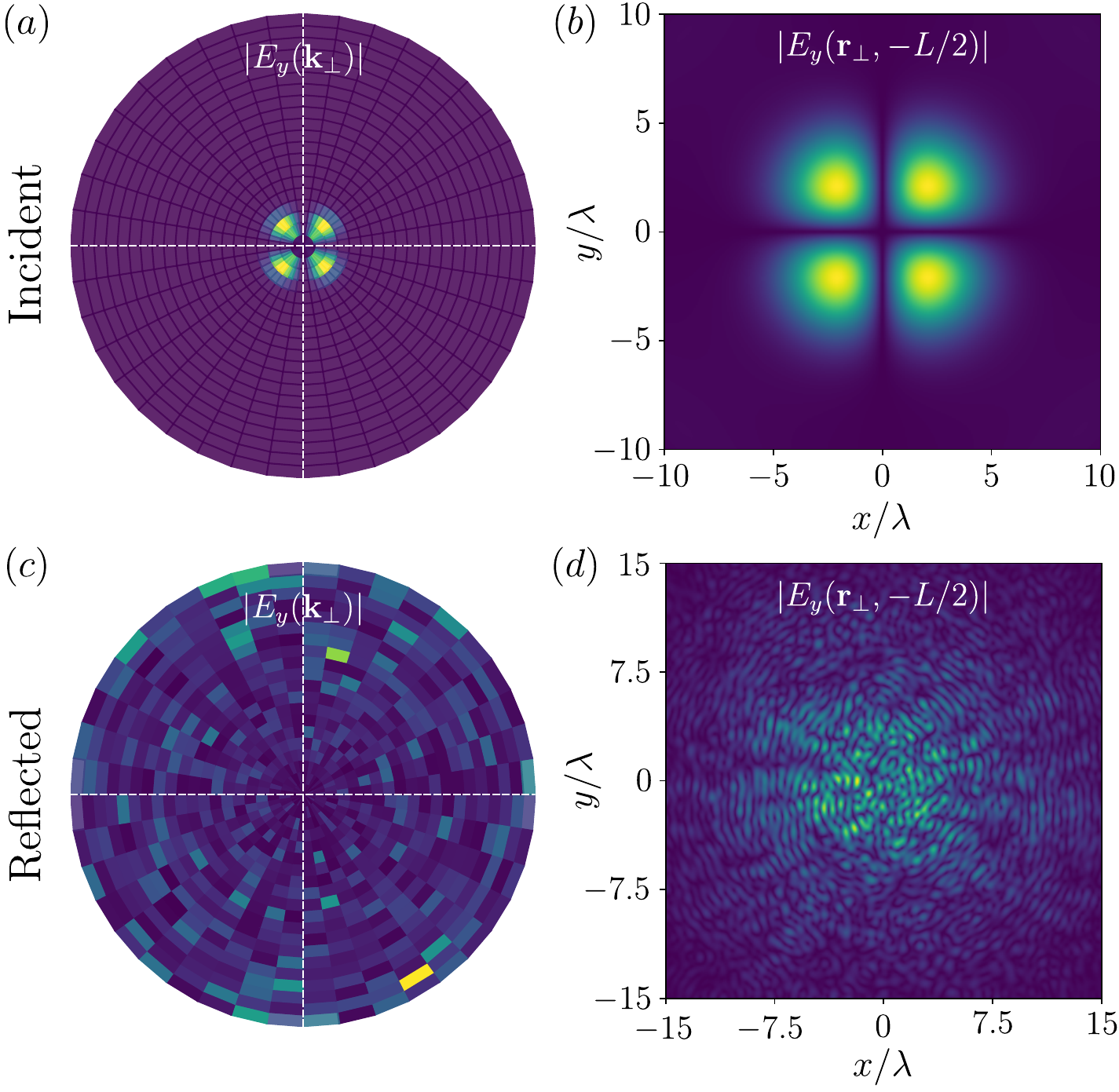} 
    \caption{Incident and reflected fields for Hermite--Gaussian illumination. The top row, (a) and (b), shows the incident field, while the bottom row, (c) and (d), shows the reflected field. The first column, (a) and (c), shows the magnitude of the $y$-component of the Fourier-space representation of the field, while the second column, (b) and (d), shows the magnitude of the $y$-component of the field at the front face of the scattering medium. Each plot uses an independent color map in arbitrary units.}
    \label{fig:fields_rayleigh}
\end{figure}

We configure the beam such that its waist, $w_0 = 3\lambda$ with $\lambda = 500$\,nm, lies at the front face of a thin scattering medium, i.e., at $z=-L/2$, where the medium thickness $L$ is much smaller than the mean free path, and calculate its angular spectrum using Eq.~(\ref{eq:gaussian-beam}). Fig.~\ref{fig:fields_rayleigh}(a) shows the magnitude of the $y$-component of the Fourier-space representation of the incident electric field at the entrance face. In particular, we present averaged field values calculated over each region according to Eq.~(\ref{eq:discretea}). Our choice of beam parameters corresponds to a relatively tight focus at the front face of the medium, and thus a relatively broad distribution in Fourier space. Fig.~\ref{fig:fields_rayleigh}(b) shows the magnitude of the $y$-component of the electric field in the plane of the medium's front face, calculated using Eq.~(\ref{eq:discrete}) without the transmission operator. The expected four-lobe field profile is observed.

We next generate a single random scattering matrix, which allows us to compute the field reflected by a single realization of the medium. Fig.~\ref{fig:fields_rayleigh}(c) shows the magnitude of the $y$-component of the Fourier-space representation of the \emph{reflected} field at the medium's front face, calculated directly from the scattering equation $\mathbf{O} = \mathbf{S}\mathbf{I}$. The field contains only scattered components, and the structure of the incident beam is lost. Since the medium scatters isotropically, the Fourier-space representation is relatively uniform, with random fluctuations. Fig.~\ref{fig:fields_rayleigh}(d) shows the corresponding field in real space, which exhibits a random speckle pattern contained within a weak circular envelope. Because the medium is very thin, the reflected field is much weaker than the incident field. Each subplot of Fig.~\ref{fig:fields_rayleigh} therefore uses an independent color map in arbitrary units, and direct comparisons between subplots should not be made. 

\subsection{Memory effect}\label{sec:memory}

\begin{figure*}[t] \includegraphics[width=\textwidth]{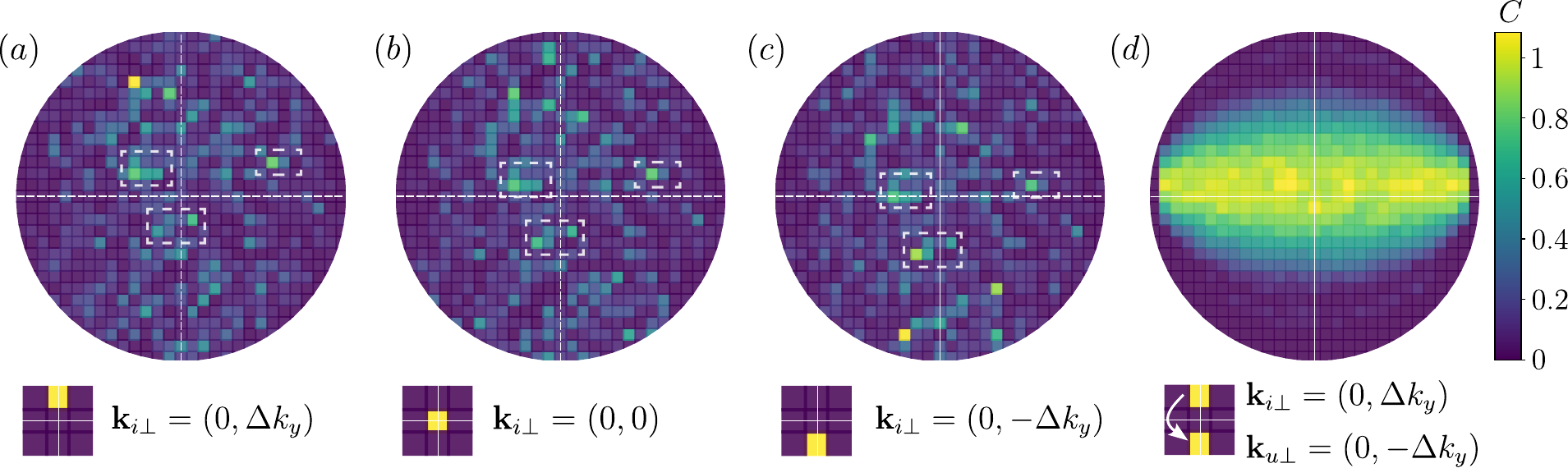} 
    \caption{(a)--(c): Fourier-space intensity of the field reflected by a thin scattering medium under illumination by three $x$-polarized plane waves with distinct incident wavevectors. Prominent features that shift in accordance with the memory effect are highlighted by dashed white boxes. (d): \red{Normalized Fourier-space correlation function associated with an incident plane wave tilt of $-2\Delta k_y$.}}
    \label{fig:memory-effect-tilts}
\end{figure*}

We now demonstrate the memory effect in our simulations. Although the memory effect is fundamentally a statistical phenomenon, it is known to manifest even after averaging only a small number of realizations~\cite{Bertolotti2012} and, at least for thin scattering media, is apparent in our simulated data at the level of single realizations. We show in particular that our simulated speckle patterns exhibit the traditional tilt--tilt memory effect, as described by Eq.~(\ref{eq:mem-eff}). We leave the analysis of more general memory effects, such as shift--shift correlations~\cite{Osnabrugge:17}, for future work.

To better visualize the memory effect, we now employ a square lattice partition, such as that shown in Fig.~\ref{fig:partitions}(a). Lattice partitions are well suited to this task because their translational symmetry allows the incident field to be tilted in evenly spaced increments of $\mathbf{k}_\perp$. Moreover, as discussed in Sec.~\ref{sec:corr}, lattice partitions admit additional simplifications that allow for faster computation of the large number of required correlations, thereby enabling an increase in the number of extended channels and an improvement in field resolution in Fourier space. In particular, we use a grid spacing of $0.07k$, which yielded a grid containing  $697$ regions. \red{We only include memory effect correlations corresponding to region quadruples whose associated dual lattice radius is 0.}

We simulate plane wave illumination by assigning nonzero amplitude to only a single region in the Fourier representation of the incident field. The center of the non-zero region is then representative of the plane wave’s wavevector. While these fields are not true plane waves, which would have zero extent in $k$-space, they are sufficient for our purposes. 

Figs.~\ref{fig:memory-effect-tilts}(a)--(c) show the transmitted Fourier-space intensity at the back plane of a thin scattering medium (again $L \ll l$) for a collection of $x$-polarized incident plane waves. In this case, the scattering medium contains spherical scatterers with size parameter equal to $2$. Fig.~\ref{fig:memory-effect-tilts}(b) corresponds to illumination for which only the central region of the partition has nonzero intensity, representing a normally incident plane wave. Figs.~\ref{fig:memory-effect-tilts}(a) and (c), by contrast, correspond to cases in which only the regions one lattice cell above and below the central region have non-zero intensity, respectively. These cases therefore correspond to incident plane waves tilted upward and downward by one lattice spacing relative to normal incidence. It is clear that the angular resolution is ultimately limited by the grid spacing, or equivalently, by the number of channels. 

Though the speckle patterns in Figs.~\ref{fig:memory-effect-tilts}(a)--(c) are random, we can clearly see features shifting in accordance with the tilt of the incident field, indicating the existence of memory effect correlations. Some more prominent shifting features are highlighted with dashed white boxes. We note that the effect diminishes near the edges of Fourier space. We attribute this to the decaying nature of the sinc function in Eq.~(\ref{eq:long2}), once the appropriate values of $k_z$ are taken into account.

\red{A more quantitative demonstration of the memory effect is shown in Fig.~\ref{fig:memory-effect-tilts}(d). We consider two transmitted Fourier-space intensity speckle patterns, $I_j$ and $I_v$, produced by incident plane waves with wavevectors $\mathbf{k}_{j\perp}$ and $\mathbf{k}_{v\perp}$, respectively, located one lattice region above and below normal incidence, as illustrated in Figs.~\ref{fig:memory-effect-tilts}(a) and (c). To align the transmitted speckle patterns, $I_v$ is shifted upward by two lattice regions to obtain $I_v'$. Values shifted outside the lattice interior are discarded, while regions at the bottom of $I_v'$ that are left undefined by the shift are assigned a value of zero. We then compute the region-wise normalized correlation function}
\begin{align}\label{eq:Icorr}
\red{C = \frac{\langle I_j I_v' \rangle}{\langle I_j \rangle \langle I_v' \rangle} - 1},
\end{align}
\red{where the averages are taken over 3000 independent realizations of the scattering matrix. As can be seen, $C$ decays in the vertical direction, but is maintained over a horizontal band where $C\sim 1$. The form of the correlation function has a $\mathrm{sinc}^2$ dependence that ultimately derives from Eq.~(\ref{eq:long2}), with the horizontal band corresponding to particular wavevector combinations for which the argument of the $\mathrm{sinc}$ vanishes. The anisotropy of $C$ has been studied in more detail in a recent work \cite{Byrnes2026b}.}

\begin{figure*}[t] \includegraphics[width=\textwidth]{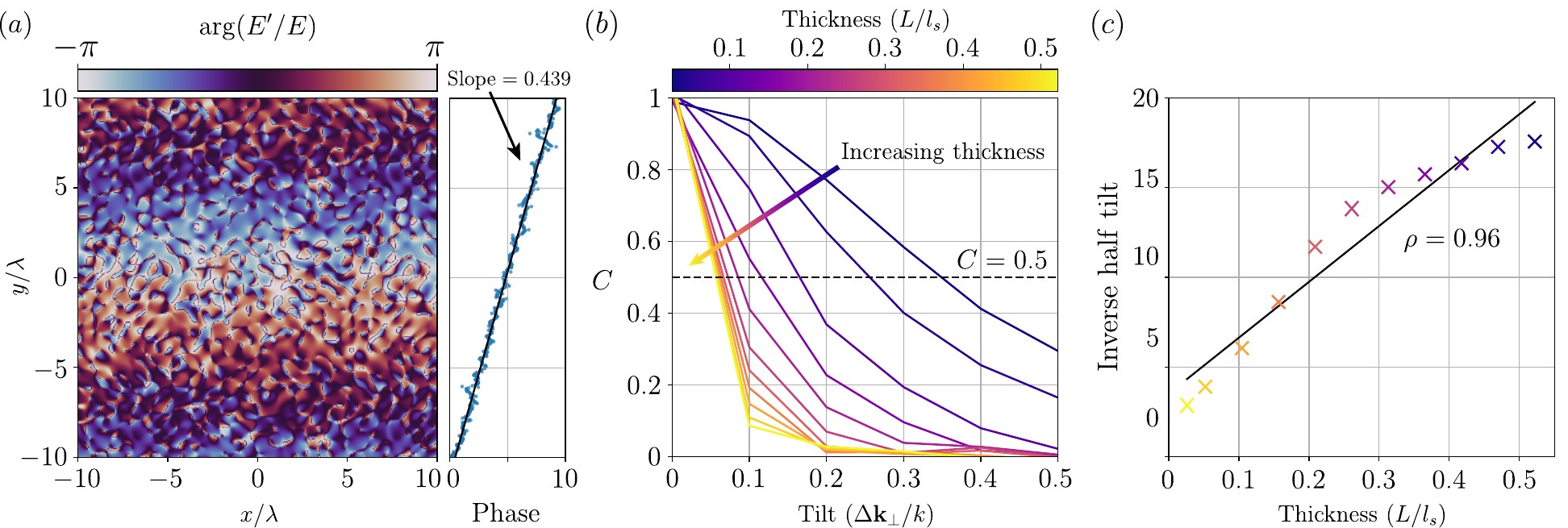} 
    \caption{\red{(a): Argument of the ratio of the transmitted field due to a tilted and untitled incident plane wave. Side panel shows the unwrapped phase of row-averaged fields for different values of $y$ presented in the main panel. (b): Plot of memory effect decay versus tilt angle for different scattering medium thicknesses. (c): Inverse values of tilts for which $C=0.5$ in (b) for different scattering medium thicknesses. The linear trend supports an inverse relation between memory effect range and medium thickness.} }
    \label{fig:memory-effect-second}
\end{figure*}

The memory effect also manifests in the real space form of the electric field. Although the speckles do not shift as they do in Fourier space, a tilt of $\Delta k_y$ in the incident wavevector imparts a phase ramp of $\exp(i\Delta k_y y)$ to the electric field. This is visible in the argument of the ratio of the electric fields for tilted and normally incident plane waves, as shown in \red{Fig.~\ref{fig:memory-effect-second}(a)}. In particular, we show the argument of the ratio of the electric fields for a vertical tilt of one lattice region from the origin. Since the lattice spacing is $0.07k$, we expect a full gradient of $2\pi$ radians over a vertical spatial extent of $\lambda/0.07 \sim 15\lambda$, which is visible within Fig.~\ref{fig:memory-effect-second}(a). \red{To demonstrate this numerically, we also plot the vertical variation of the electric field's phase in the side panel of Fig.~\ref{fig:memory-effect-second}(a). Data points show the unwrapped argument of the row-averaged electric field for each value of $y$ spanned in Fig.~\ref{fig:memory-effect-second}(a). The data are well-fit by a line of slope $0.439$, which correctly corresponds to a wavevector tilt of $0.439k/2\pi \sim 0.07k$.}

\red{As a final demonstration of the memory effect, we examine how the memory effect range depends on the thickness of the scattering medium. For this study, we consider a lattice with spacing $0.1k$ and scattering media with thicknesses up to $0.5l$. For each thickness, we compute the correlation function $C$ defined in Eq.~(\ref{eq:Icorr}), now as a function of tilt angle rather than Fourier-space region. We apply vertical plane-wave tilts of up to five lattice regions, corresponding to a maximum tilt angle of $30^\circ$. Using the transmitted speckle pattern for normal incidence, $I_j$, as a reference, we construct appropriately shifted intensity patterns $I_v'$ for each tilt angle. The correlation function $C$ is then evaluated by averaging over a subset of Fourier-space regions, allowing spatial variations in the correlation strength to be resolved. As shown in Fig.~\ref{fig:memory-effect-tilts}(d), the correlation function is anisotropic, such that we choose to average over three regions vertically aligned above the origin and three regions vertically aligned below the origin. We exclude the origin as it contains the incident field, which dominates the total field for the thicknesses considered. This choice emphasizes the $\mathrm{sinc}$-like decay of the correlation function with increasing tilt angle. The resulting spatial average is subsequently averaged over 2000 independent realizations of the scattering medium.}

\red{Our results are shown in Fig.~\ref{fig:memory-effect-second}(b). As expected, the correlation function $C$ decreases with increasing tilt angle, and the rate of decay increases with the thickness of the scattering medium. To quantify the memory effect range, we determine, for each thickness, the tilt angle at which $C=0.5$. For thick scattering media, the memory effect range is known to scale inversely with medium thickness \cite{RevModPhys.89.015005}. This scaling follows from the inverse $\mathrm{sinhc}^2$ form factor that arises in conventional ladder-approximation treatments of the memory effect. Although our study is restricted to relatively thin media, we note that the $\mathrm{sinc}^2$ and inverse $\mathrm{sinhc}^2$ functions agree to second order in their Taylor expansions, suggesting similar mathematical behavior at small tilt angles. To test for the expected scaling, we plot the inverse of the tilt angle satisfying $C=0.5$ as a function of medium thickness in Fig.~\ref{fig:memory-effect-second}(c). These values are found using linear interpolation between the data points in Fig.~\ref{fig:memory-effect-second}(b). Under the inverse-scaling relation, the data should lie on a straight line. A linear fit shows good agreement with this prediction, yielding a Pearson correlation coefficient of $\rho = 0.96$. We attribute deviations from the linear fit, particularly at intermediary thicknesses, to insufficient angular resolution in our data, which is ultimately tied to our region size.}

\red{Ideally, we would like to investigate the statistical properties of the memory effect deep within the multiple-scattering regime. In practice, however, the memory effect range rapidly decreases with increasing medium thickness and soon falls below the angular resolution provided by the lattice discretizations used in the present demonstrations. In principle, this limitation can be mitigated by employing finer lattice partitions. Such refinements, however, lead to correspondingly larger scattering matrices and increased computational cost. We note that the flexibility of the partitioning may permit the design of bespoke lattices for such studies, for example by sampling Fourier space finely along one direction and more coarsely along another. The exploration of such strategies, however, lies beyond the scope of the present work.}

\subsection{Propagation through thick media}\label{sec:cascade}

We now consider the transmission and reflection of a vectorial beam through scattering media of varying thicknesses. We use a polar partition with $\Delta r = 0.1k$ and $\Delta \theta = 2\pi/20$, giving a total of $200$ channels. \red{To simplify our simulations, we `turn off' the memory effect by again only calculating auto-correlations for scattering matrix sub-blocks.} For the scattering medium, we choose spherical scatterers with size parameter $2$ and relative refractive index $1.2$. We now take $L = 1.126\,\mu\mathrm{m}$ and $n = 0.592\,\mu\mathrm{m}^{-3} $, which corresponds to a mean free path of $l \sim 88\,\mu m$. Spheres of this size scatter primarily in the forward direction, with an anisotropy factor $g \sim 0.66$, giving a transport mean free path of $l_t = l/(1-g) \sim 260\,\mu m$. For our incident field, we construct Gaussian beams, which, in the waist plane, are assumed to have the generic form
\begin{align}
    \mathbf{E}(k_x,k_y) = 
    e^{-\frac{w_0^2}{4}(k_x^2+k_y^2)}
    \left(\mathbf{p} - [\mathbf{p}\cdot\widehat{\mathbf{e}}_k]\widehat{\mathbf{e}}_k\right),
\end{align}
where now $w_0 = 5\lambda$.  We consider three different types of polarization structures for our beam, corresponding to three different choices for the function $\mathbf{p}$: 
\begin{align}
    \widehat{\mathbf{p}}(\mathbf
    {k}_\perp) = 
    \begin{cases}
        \widehat{\mathbf{y}} & \text{linear} \\
        (\widehat{\mathbf{x}} + i \widehat{\mathbf{y}})/\sqrt{2}  & \text{circular}\\
        -k_y\widehat{\mathbf{x}} + k_x\widehat{\mathbf{y}} & \text{azimuthal} 
\end{cases}.
\end{align}
The linear and circular polarizations, aside from projection effects, are uniform across the beam, whereas the azimuthal polarization corresponds to a rotating linear polarization, giving rise to a polarization singularity at $\mathbf{k}_\perp = \mathbf{0}$. Through cascading, we construct scattering matrices corresponding to media of thicknesses ranging from $0 \leq L/l_t \leq 8$, collecting data from $700$ different realizations of the scattering medium.

At each thickness where data is collected, and for each beam polarization structure, we compute three quantities in both transmission and reflection: one measure of average scattered intensity and two measures of degree of polarization (DoP). These can be obtained from the Stokes vectors of each scattering channel. Since the two orthogonal field components of a channel form a Jones vector, the corresponding Stokes vector $(S_0,S_1,S_2,S_3)^\mathrm{T}$ can be calculated directly~\cite{Gil2022}.

\begin{figure}[t] 
    \centering    \includegraphics[width=\columnwidth]{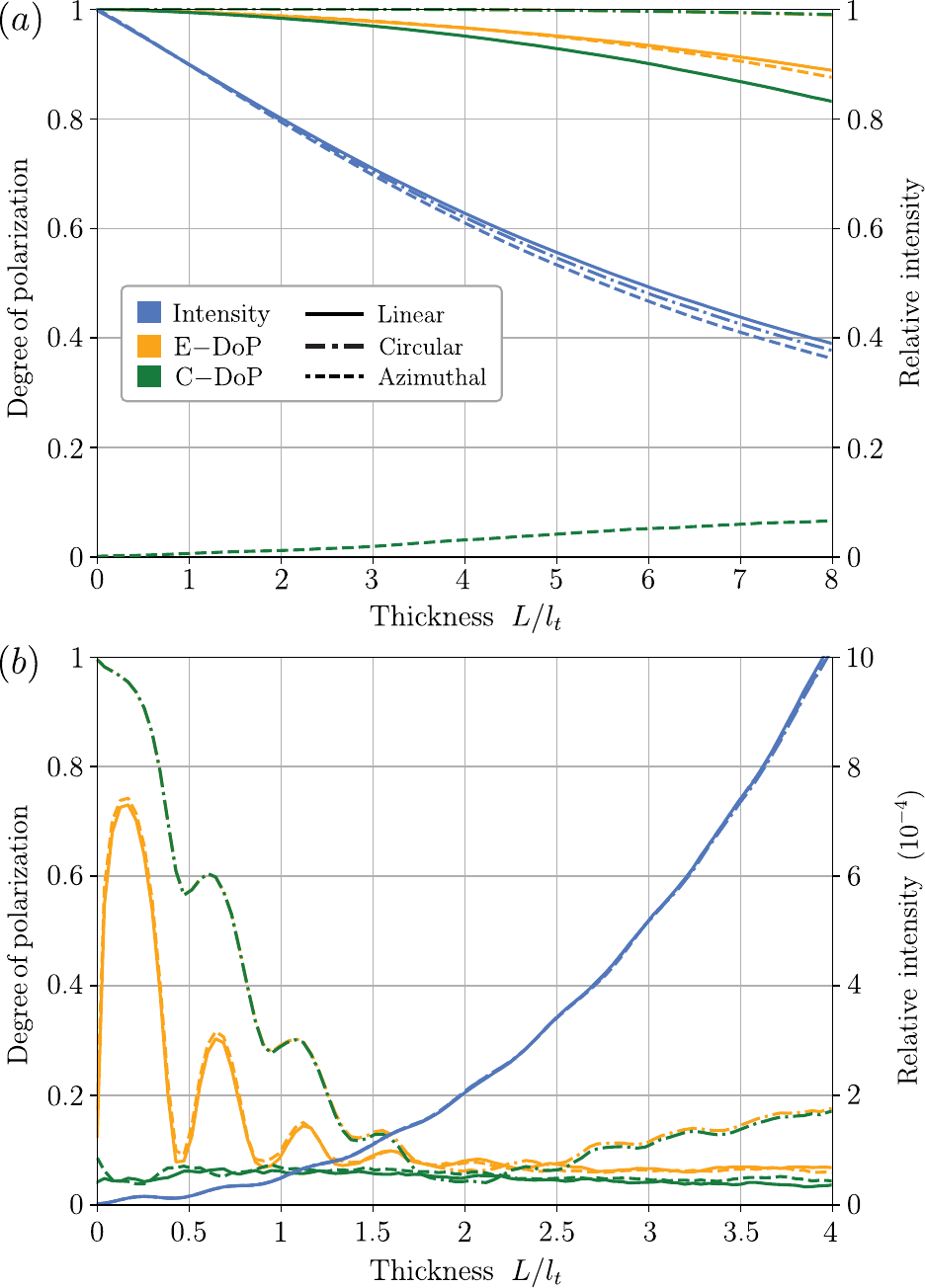} 
    \caption{Mean intensity (blue), E-DoP (orange), and C-DoP (green) versus medium thickness calculated in transmission (a) and reflection (b) for incident Gaussian beams with linear (solid), circular (dot-dashed), and azimuthal (dashed) polarization structures.}
    \label{fig:cascade}
\end{figure}

For each random realization, we find $S_0$ for the Stokes vector averaged over the central ring of channels in the partition, i.e., the ring of regions with radial extent $0 \leq r \leq \Delta r$. This provides a per-realization measure of the transmitted (reflected) intensity along (opposite) the incident beam's dominant propagation direction. Dividing these values by the ring-averaged $S_0$ values for the incident beams then gives us measures of the relative transmitted and reflected intensities. Finally, we average these values over the ensemble of random medium realizations. These ensemble averaged relative intensities are are shown as the blue curves in Figs.~\ref{fig:cascade}(a) and (b), respectively with different incident polarization states shown through different line styles. The general trends are clear and independent of the incident polarization: transmitted intensity decreases with thickness, while reflected intensity increases, as expected from energy conservation.

For any given Stokes vector, its DoP is given by $\mathrm{DoP} = \sqrt{S_1^2 + S_2^2 + S_3^2}/S_0$. We consider two distinct applications of this formula. First, for each channel in the central ring, we find the DoP of the ensemble averaged Stokes vector. We then average these values over the channels in the ring. We refer to this final quantity as the \emph{ensemble} DoP (E-DoP). In the second approach, we first compute the DoP of the ring-averaged Stokes vector for each realization. This value is then averaged over all realizations. We refer to this final quantity as the \emph{channel} DoP (C-DoP). Roughly, E-DoP captures global features across the ensemble of random media, while C-DoP reveals variations in polarization over the extent of the beam. Figs.~\ref{fig:cascade}(a) and (b)  also show both measures of DoP against medium thickness in transmission and reflection. In transmission, E-DoP for all three beams decays with thickness. The rates, however, depend on the incident field's polarization structure. While depolarization rates are similar for the linear and azimuthal beams, polarization is much more strongly preserved for the circular beam. This is a manifestation of the well known polarization memory effect \cite{Xu2005} (not to be confused with the memory effect discussed throughout this paper).

For the linear and circular beams, C-DoP behaves similarly to E-DoP. For the azimuthal beam, however, the C-DoP increases with thickness, starting from a value of 0. The initial vanishing of C-DoP is a consequence of the rotational nature of the azimuthal polarization, for which the ring-averaged Stokes vector is unpolarized by construction. As the beam propagates through the medium, however, the linear polarization states around the central ring progressively randomize, breaking this symmetry and resulting in partial polarization when averaging over the central ring.

In reflection, the degree of polarization exhibits more complex behavior. For example, E-DoP for the linear and azimuthal beams is oscillatory with a period of roughly $0.5\,l_t$ and a decaying envelope. This can be understood by considering the reflected field as consisting of two components: the mean field, which is fully polarized and a random background that tends to depolarize the field. For thin media, the mean field is stronger and dominates the scattering response. As indicated by Eq.~(\ref{eq:t-av}), the mean reflection matrix is initially oscillatory. At thicknesses where the mean reflection is large, E-DoP is therefore high, while it is small at thicknesses where the mean reflection is low. The decaying envelope arises as the medium becomes thicker and random scattering begins to dominate. For C-DoP, the same underlying physics applies; however the degree of polarization remains small even at small thicknesses due to averaging over the central ring.

For the circular beam, a different mechanism, initially described in Ref.~\cite{Byrnes14122022}, governs the behavior. At small thicknesses, both components of the field consist primarily of circularly polarized light with opposite chirality to the incident beam. At larger thicknesses, due to the polarization memory effect, the random background tends to preserve the incident beam's polarization state. Consequently, the DoP is high for thin media and remains notably nonzero for thick media, where the random background begins to dominate. We note that the point where DoP begins to grow for the circular beam, at approximately $2$ to $2.5\,l_t$, coincides with the thickness where the E-DoP for the linear and azimuthal beams has fully decayed, marking a transition to a regime of diffuse scattering.

\section{Conclusion}
In this work, we have presented an enhanced random matrix simulation framework for modeling the propagation of polarized light through random scattering media composed of scatterers with tailored physical properties. Through the introduction of extended scattering channels, the framework supports a broad range of simulation scenarios, including diverse beam profiles, memory effect configurations, and improved alignment with specific detector geometries. It further enables a more rigorous treatment of scattering matrix correlations and offers new geometric insights. In addition, we have provided a freely available supporting codebase containing a flexible and expandable implementation of the proposed approach~\cite{rmtnew}.

We anticipate that this framework will facilitate a wide range of statistical studies in random systems, including beam propagation through disordered media, polarization effects and depolarization analysis, memory-effect investigations, speckle correlation analysis, and the study of scattering eigenchannels. Beyond these applications, the framework is also well suited for investigations of wavefront shaping, imaging through scattering media, mesoscopic transport phenomena, random-laser behavior, and information transport in disordered optical systems.

\begin{acknowledgments}
N.B. and M.R.F. were supported by the Nanyang Technological University Grant SUG:022824-00001. M.R.F. acknowledges additional funding from the Institute for Digital Molecular Analytics and Science (IDMxS) under the Singapore Ministry of Education Research Centres of Excellence scheme (EDUN C-33-18-279-V12). S.D. was funded by an Interdisciplinary Graduate Programme PhD Research Scholarship through IDMxS.
\end{acknowledgments}

\section*{Author Contributions}
\textbf{Niall Byrnes}: Methodology, Software, Validation, Formal analysis, Investigation, Data Curation, Writing - Original Draft, Visualization.
\textbf{Sulagna Dutta}: Validation, Investigation, Visualization, Writing - Review \& Editing.
\textbf{Matthew R. Foreman}: Conceptualization, Validation, Resources, Writing - Review \& Editing, Visualization, Supervision, Project administration, Funding acquisition.

\appendix

\section{Scattering matrix symmetries}
In this section we examine symmetry constraints for the scattering matrix defined with respect to extended scattering channels. Of particular interest are energy conservation and reciprocity.

\subsection{Energy conservation}\label{sec:energy}
In this section we discuss energy conservation and explain the renormalization of the scattering matrix as introduced in Eq.~(\ref{eq:sdisc}).

Following Ref.~\cite{Byrnes2021a} and using the partition discussed in the main text, energy conservation for the incident and outgoing fields $\mathbf{I}(\mathbf{k}_\perp)$ and $\mathbf{O}(\mathbf{k}_\perp)$ can be expressed as
\begin{align}\label{eq:en-cons}
    \sum_{i=1}^N\int_{K_i} \bigg(|\mathbf{I}(\mathbf{k}_\perp)|^2 - |\mathbf{O}(\mathbf{k}_\perp)|^2\bigg)\,\dee \mathbf{k}_\perp = 0.
\end{align}
Applying the same discrete approximation discussed in the main text, Eq.~(\ref{eq:en-cons}) can be written in matrix form, viz.,
\begin{align}\label{eq:en-cons2}
    \mathbf{I}^\dagger\mathbf{W}\mathbf{I} - \mathbf{O}^\dagger\mathbf{W}\mathbf{O} = \mathbb{O},
\end{align}
where $\mathbf{I}$ and $\mathbf{O}$ are given by Eqs.~(\ref{eq:discI}) and (\ref{eq:discO}), $\mathbf{W} = \mathbb{I}_2 \otimes\mathrm{diag}(w_1, w_2, \dots,w_N) \otimes \mathbb{I}_2$, $\mathbb{O}$ is the zero matrix, $\mathbb{I}_2$ is the $2\times 2$ identity matrix, and $\otimes$ is the Kronecker product. Since $\mathbf{O} = \mathbf{S}\mathbf{I}$, Eq.~(\ref{eq:en-cons2}) can be written as $\mathbf{I}^\dagger(\mathbf{W} - \mathbf{S}^\dagger\mathbf{W}\mathbf{S})\mathbf{I} = \mathbb{O}$. This holds for all $\mathbf{I}$, implying
\begin{align}\label{eq:en-cons3}
    \mathbf{S}^\dagger\mathbf{W}\mathbf{S} = \mathbf{W}.
\end{align}
Evidently, the presence of the weight matrix $\mathbf{W}$ means energy conservation does not correspond to unitarity for our particular definition of $\mathbf{S}$. Unitarity can be recovered, however, by defining the renormalized scattering matrix 
\begin{align}\label{eq:renorm}
    \widetilde{\mathbf{S}} = \sqrt{\mathbf{W}}\mathbf{S}\sqrt{\mathbf{W}}^{-1},
\end{align}
with which it is straightforward to show that Eq.~(\ref{eq:en-cons3}) can be reduced to the unitarity condition $\widetilde{\mathbf{S}}^\dagger\widetilde{\mathbf{S}} = \mathbb{I}$. Careful inspection of Eq.~(\ref{eq:renorm}) shows that $\widetilde{\mathbf{S}}$ is related to $\mathbf{S}$ in precisely the sense described in the main text, i.e., in the sense that sub-blocks satisfy $\widetilde{\mathbf{s}}_{(j,i)} = \sqrt{w_j / w_i}\mathbf{s}_{(j,i)}$.

\subsection{Reciprocity}
\label{sec:reciprocity}
For an inversion-symmetric partition with numbering scheme described in the main text, every sub-block $\widetilde{\mathbf{s}}_{(j,i)}$ has a reciprocal partner with which it shares a simple, deterministic relationship when the underlying scattering medium obeys reciprocity. This follows straightforwardly from Eq.~(\ref{eq:sdisc}) and known reciprocity relations for the continuous scattering matrix~\cite{Byrnes2021a}.

Reciprocity for the continuous scattering matrix is given by
\begin{align}
    \mathbf{r}(\mathbf{k}_{j\perp}, \mathbf{k}_{i\perp}) &=     \mathbf{r}^\mathrm{R}(-\mathbf{k}_{i\perp}, -\mathbf{k}_{j\perp}),\label{eq:rec1}\\
    \mathbf{r}'(\mathbf{k}_{j\perp}, \mathbf{k}_{i\perp}) &=     \mathbf{r}'^{\mathrm{R}}(-\mathbf{k}_{i\perp}, -\mathbf{k}_{j\perp}),\label{eq:rec2}\\
    \mathbf{t}(\mathbf{k}_{j\perp}, \mathbf{k}_{i\perp}) &=     \mathbf{t}'{^\mathrm{R}}(-\mathbf{k}_{i\perp}, -\mathbf{k}_{j\perp}),\label{eq:rec3}
\end{align}
where $\mathrm{R}$ is a reciprocal operator, which, for $2\times 2$ matrices, is defined by 
$\mathbf{s}^\mathrm{R} = \boldsymbol{\sigma}_z\mathbf{s}^\mathrm{T}\boldsymbol{\sigma}_z$ where $\boldsymbol{\sigma}_z = \mathrm{diag}(1,-1)$ is a Pauli matrix. Using straightforward changes of variables, we find
\begin{align}
\begin{split}
    \widetilde{\mathbf{s}}_{1(j,i)} &= \frac{1}{\sqrt{w_i w_j}}\int_{K_i \times K_j}\widetilde{\mathbf{s}}_{1}(\mathbf{k}_{j\perp}, \mathbf{k}_{i\perp})\,\dee\mathbf{k}_{j\perp}\dee\mathbf{k}_{i\perp}\\
    &= \frac{1}{\sqrt{w_i w_j}}\int_{K_i \times K_j}\widetilde{\mathbf{s}}_{2}^\mathrm{R}(-\mathbf{k}_{i\perp}, -\mathbf{k}_{j\perp})\,\dee\mathbf{k}_{j\perp}\dee\mathbf{k}_{i\perp}\\
    &= \frac{1}{\sqrt{w_i w_j}}\int_{-K_j \times -K_i}\widetilde{\mathbf{s}}_{2}^\mathrm{R}(\mathbf{k}_{i\perp}, \mathbf{k}_{j\perp})\,\dee\mathbf{k}_{i\perp}\dee\mathbf{k}_{j\perp}\\
    &= \frac{1}{\sqrt{w_{-i}w_{-j}}}\int_{K_{-j} \times K_{-i}}\widetilde{\mathbf{s}}_{2}^\mathrm{R}(\mathbf{k}_{i\perp}, \mathbf{k}_{j\perp})\,\dee\mathbf{k}_{i\perp}\dee\mathbf{k}_{j\perp}\\
    &= \widetilde{\mathbf{s}}^\mathrm{R}_{2(-i,-j)}.\label{eq:recdev}
\end{split}
\end{align}
Note that in going from the first to second line in Eq.~(\ref{eq:recdev}), $\widetilde{\mathbf{s}}_{1}(\mathbf{k}_{j\perp}, \mathbf{k}_{i\perp})$ is replaced with $\widetilde{\mathbf{s}}_{2}^\mathrm{R}(-\mathbf{k}_{i\perp}, -\mathbf{k}_{j\perp})$ using Eqs.~(\ref{eq:rec1})-(\ref{eq:rec3}). The matrices $\widetilde{\mathbf{s}}_{1(j,i)}$ and $\widetilde{\mathbf{s}}^\mathrm{R}_{2(-i,-j)}$ are therefore reciprocal partners. It can further be shown that the entire scattering matrix $\widetilde{\mathbf{S}}$ satisfies the reciprocity formula~\cite{ByrnesPhD}
\begin{align}
    \widetilde{\mathbf{S}} = (\mathbb{I}_2 \otimes \mathbf{J}_{N}\otimes \boldsymbol{\sigma}_z )\widetilde{\mathbf{S}}^\mathrm{T}(\mathbb{I}_2 \otimes \mathbf{J}_{N}\otimes \boldsymbol{\sigma}_z ),
\end{align}
where $\mathbf{J}_N$ is the exchange matrix of size $N$, and $N$ is the number of regions in the partition.

\section{Minkowski filter}
\label{sec:minkowski}

In this section we present a method for estimating intersection volumes for the purpose of filtering covariance candidates for arbitrary partitions. Given the emergence of Minkowski sums, we refer to this approach as the Minkowski filter.

\begin{figure}[b]   
\includegraphics[width=\columnwidth]{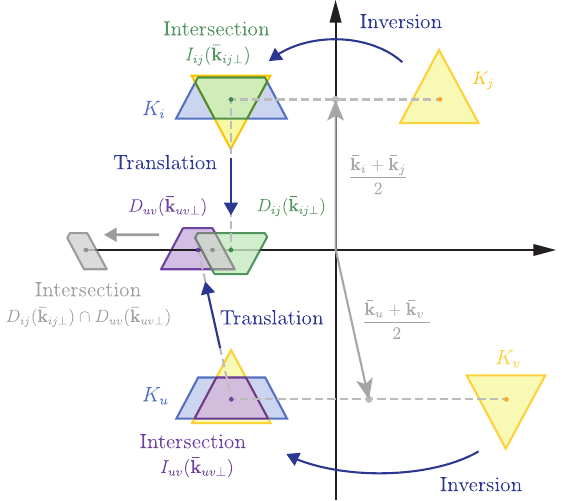} 
    \caption{Example illustrations of the Minkowski filter applied to two different quadruples of regions. The approximate upper bound for the area of $D(\mathbf{\bar{k}}_{ij\perp}) \cap D(\mathbf{\bar{k}}_{uv\perp})$ is given by the area of the gray shapes labeled `intersection' at the left sides of each sub-figure.}
    \label{fig:mink}
\end{figure}

Returning to Eqs.~(\ref{eq:long1}) and (\ref{eq:long2}), the argument of the delta function can be simplified by applying the mean-difference coordinate transformations to the pairs of variables $(\mathbf{k}_{i\perp}, \mathbf{k}_{j\perp})$ and $(\mathbf{k}_{u\perp}, \mathbf{k}_{v\perp})$. In particular, we write $\bar{\mathbf{k}}_{ij\perp} = (\mathbf{k}_{i\perp} + \mathbf{k}_{j\perp})/2$ and $\Delta \mathbf{k}_{ij\perp} = \mathbf{k}_{i\perp} - \mathbf{k}_{j\perp}$, and similarly for $\mathbf{k}_{u\perp}$ and $\mathbf{k}_{v\perp}$. The integral in Eq.~(\ref{eq:long1}) then transforms according to
\begin{align}
    \int_{K_i \times K_j \times K_u \times K_v} \to \int_{\frac{K_i + K_j}{2}}\int_{\frac{K_u + K_v}{2}}\int_{D_{ij}(\mathbf{\bar{k}}_{ij\perp}) \times D_{uv}(\mathbf{\bar{k}}_{uv\perp})},
\end{align}
where, e.g., $K_i + K_j$ is the Minkowski sum of $K_i$ and $K_j$, and $D_{ij}(\bar{\mathbf{k}}_{ij\perp}) = \{ \Delta\mathbf{k}_{ij\perp} \mid \bar{\mathbf{k}}_{ij\perp} + \Delta\mathbf{k}_{ij\perp}/2 \in K_i, \bar{\mathbf{k}}_{ij\perp} - \Delta\mathbf{k}_{ij\perp}/2 \in K_j \}$ is the set of possible differences between vectors in $K_i$ and $K_j$ for a fixed value of $\bar{\mathbf{k}}_{ij\perp}$. Under this change of variables, the delta function in Eq.~(\ref{eq:long2}) transforms to $\delta(\Delta\mathbf{k}_{ij\perp} - \Delta\mathbf{k}_{uv\perp})$. Integrating this over the final 4-dimensional domain $D_{ij}(\mathbf{\bar{k}}_{ij\perp}) \times D_{uv}(\mathbf{\bar{k}}_{uv\perp})$ causes the final domain to collapse to the 2-dimensional domain $D_{ij}(\mathbf{\bar{k}}_{ij\perp}) \cap D_{uv}(\mathbf{\bar{k}}_{uv\perp})$ and enforces the constraint $\Delta\mathbf{k}_{ij\perp} = \Delta\mathbf{k}_{uv\perp}$ within the integrand. We now seek to approximate the area of $D_{ij}(\mathbf{\bar{k}}_{ij\perp}) \cap D_{uv}(\mathbf{\bar{k}}_{uv\perp})$. 

The area of $D_{ij}(\mathbf{\bar{k}}_{ij\perp})$ depends on the choice of $\mathbf{\bar{k}}_{ij\perp}$. Given a particular value of $\mathbf{\bar{k}}_{ij\perp}$, we can find the shape of $D_{ij}(\mathbf{\bar{k}}_{ij\perp})$ by geometrically inverting the region $K_j$ through the point $\mathbf{\bar{k}}_{ij\perp}$ and intersecting the resulting shape with $K_i$ to obtain $I_{ij}(\mathbf{\bar{k}}_{ij\perp})$. This process is outlined in Fig.~{\ref{fig:mink}}. By construction, $I_{ij}(\mathbf{\bar{k}}_{ij\perp})$ is a translated and scaled version of $D(\mathbf{\bar{k}}_{ij\perp})$ with a quarter of its area. For a given choice of $\bar{\mathbf{k}}_{uv\perp}$, the same procedure can be applied to $K_u$ and $K_v$ to obtain $I_{uv}(\mathbf{\bar{k}}_{uv\perp})$. To find $D(\mathbf{\bar{k}}_{ij\perp}) \cap D(\mathbf{\bar{k}}_{uv\perp})$, we translate $I_{ij}(\mathbf{\bar{k}}_{ij\perp})$ and $I_{uv}(\mathbf{\bar{k}}_{uv\perp})$ by $-\bar{\mathbf{k}}_{ij\perp}$ and $-\bar{\mathbf{k}}_{uv\perp}$ respectively and scale the resulting region by a scale factor of $2$ about the origin. Critically, we expect $I_{ij}(\mathbf{\bar{k}}_{ij\perp})$ to be relatively large when $\mathbf{k}_{i\perp}$ and $\mathbf{k}_{j\perp}$ are close to the centers of $K_i$ and $K_j$, implying that $\bar{\mathbf{k}}_{ij\perp}$ is close to the center of $(K_i + K_j)/2$. We therefore expect the area of $D(\mathbf{\bar{k}}_{ij\perp}) \cap D(\mathbf{\bar{k}}_{uv\perp})$ to be relatively large when $\bar{\mathbf{k}}_{ij\perp}$ and $\bar{\mathbf{k}}_{uv\perp}$ are formed from the centers of the four regions. We can use the area of  $D(\mathbf{\bar{k}}_{ij\perp}) \cap D(\mathbf{\bar{k}}_{uv\perp})$
in this particular case as an estimate for the upper bound of $D(\mathbf{\bar{k}}_{ij\perp}) \cap D(\mathbf{\bar{k}}_{uv\perp})$ for all possible choices of means. The product of this upper bound and the areas of the Minkowski means gives a well-defined 6-dimensional volume. Though clearly not equivalent to the actual 6-dimensional volume $\sigma$ discussed in the main text, this heuristic volume is fast to calculate and can also be used as a filtering mechanism when compared against a suitably chosen threshold value.

%


\end{document}